\documentclass[english,prd,superscriptaddress,nofootinbib,preprintnumbers,showpacs]{revtex4}
\usepackage[latin1]{inputenc}
\usepackage{graphicx}
\usepackage{amsmath,amsthm,amssymb}
\usepackage{bm}

\usepackage{amsfonts}
\usepackage{dcolumn}

\usepackage{color}

\newcommand{\E}{{\cal E}}

\newcommand{\beqa}{\begin{eqnarray}}
\newcommand{\eeqa}{\end{eqnarray}}

\usepackage{babel}
\begin{document}

\title{Motion of Charged Particles around a Weakly Magnetized Rotating Black Hole
}

\author{Ryo Shiose
}
\affiliation{Department of Physics, College of Humanities and Sciences, 
Nihon University, Tokyo 156-8550, Japan}
\author{Masashi Kimura
}
\affiliation{DAMTP, University of Cambridge, Centre for Mathematical Sciences,
Wilberforce Road, Cambridge CB3 0WA, UK}
\author{Takeshi Chiba
}
\affiliation{Department of Physics, College of Humanities and Sciences, 
Nihon University, Tokyo 156-8550, Japan}

\begin{abstract}
We study the motion of a charged particle around a weakly magnetized rotating black hole. 
We classify the fate of a 
charged particle kicked out from the innermost stable 
circular orbit.  
We find that the final fate of the charged particle depends mostly on 
the energy of the particle and the radius of the orbit.  
The energy and the radius in turn
 depend on the initial velocity, the black hole spin,  
and the magnitude of the magnetic field. 
We also find possible evidence for the existence of bound motion 
in the vicinity of the equatorial plane.  
\end{abstract}

\date{\today}


\maketitle

\section{Introduction}

Black holes (BHs) are ubiquitous in the Universe and play an 
important role in 
the formation of galaxies~\cite{rees}.  BHs produce intense radiation 
by converting the gravitational binding energy of accreting plasmas~\cite{ss}.   
Accreting BHs immersed in large-scale  
 magnetic fields also release 
their rotational energy  into powerful relativistic jets~\cite{bz}, 
which are observed in active galactic nuclei, quasars, or X-ray binaries. 
Numerical simulations demonstrated that powerful jets are generated by extracting 
energy from a spinning BH along the magnetic field~\cite{tnm}. 

Moreover, the power of jets from BHs with thick accretion disks depends mostly on BH spin, 
which may explain the wide variety of radio luminosities of active galactic nuclei~\cite{tnm2}. 
BH spin is measured by the X-ray reflection method~\cite{fabian} (see~\cite{reynolds} for a recent review) and by 
the continuum-fitting method~\cite{zcc} (see~\cite{mn} for a recent review), and 
it is found that a large fraction of astrophysical 
 BHs are rapidly rotating. 

In this paper, in order to examine the effects of a magnetic field and BH spin 
on 
particle motion,  we investigate the motion of a charged particle around 
a rotating black hole in a uniform magnetic field.   
Although this 
 is a simplified problem, the dynamics are  still 
nonintegrable due to the lack of a 
third constant of motion 
(the Carter constant) in the presence of a magnetic field. 
If we focus on 
equatorial motion,
a semi-analytical approach is possible~\cite{Prasanna:1978vh, aliev, igata},
but 
a general orbit requires numerics~\cite{Preti:2009zz, Preti:2010zza, fs, zfs, Hussain:2014cba}. 
More specifically, 
we consider the motion of a charged particle kicked out from 
the equatorial plane, and investigate 
the conditions under which such a particle can escape to infinity. 
This 
problem was studied in~\cite{Preti:2009zz, fs, zfs,frolov} for a nonrotating black hole
and in~\cite{Hussain:2014cba} for a slowly rotating black hole.
There, it was  
found that the charged particle is 
either captured by the black hole, or escapes in a direction parallel
or antiparallel to the magnetic field. The final fate of this particle  
is extremely sensitive to the initial conditions determined by the strength of 
the magnetic field. 
Since astrophysical BHs are rotating (sometimes rapidly),
we extend such an analysis to a rotating black hole. 

The paper is organized as follows. In Sec.~\ref{sec2}, after introducing 
a  
uniform magnetic field, we present the equations of motion for 
a charged 
particle and study the innermost stable circular orbits. 
In Sec.~\ref{sec3}, we present the results of numerical calculations for  
particles kicked 
out of 
these 
innermost stable circular orbits and discuss the final fate of the particles. 
We summarize our results in Sec.~\ref{secsum}.

Appendix~\ref{appendixa} contains an analysis of the magnetic flux 
across a black hole for two field configurations. 
In Appendix~\ref{appendixb}, we present approximate solutions for 
the innermost stable circular orbits.
We use 
units in which 
$G=c=1$.

Note added: during the preparation of our paper, we became aware of 
a work on similar topics~\cite{zahrani}. While it has some overlaps with our paper,
the analysis  of motion of a particle kicked out from a circular orbit in~\cite{zahrani} is limited to a fixed value of a magnetic field.
Our analysis is complementary to the results of~\cite{zahrani}.

\section{Weakly Magnetized Rotating Black Hole}
\label{sec2}

We consider the motion of charged particles in a weakly magnetized rotating black hole.
By ``weakly magnetized,'' we mean that the energy density of the magnetic field does not significantly distort the background black hole geometry which is assumed to be given by the Kerr metric (in the Boyer-Lindquist coordinates),
\beqa
ds^2&=&g_{\mu\nu}dx^{\mu}dx^{\nu}\nonumber\\
&=&
-\left(1-\frac{2M}{\Sigma}\right)dt^2-
\frac{4aMr\sin^2\theta}{\Sigma}dtd\phi+\frac{\Sigma}{\Delta}dr^2+\Sigma d\theta^2
+\frac{(r^2+a^2)^2-a^2\Delta\sin^2\theta}{\Sigma}\sin^2\theta d\phi^2
\label{kerr}
\eeqa
where $M$ is the gravitational mass of the black hole and $a$ is 
its angular momentum per unit mass 
and $\Sigma=r^2+a^2\cos^2\theta$ and $\Delta=r^2+a^2-2Mr$.  The event horizon is located at $r_H=M+\sqrt{M^2-a^2}$. 

The effect of the magnetic field on the background geometry can be neglected if 
\beqa 
GB^2\ll (GM)^{-2}   \quad{\rm or}  \quad B\ll G^{-3/2}M^{-1}\sim 10^{19} {\rm Gauss}(M_{\odot}/M),
\label{criticalB}
\eeqa
where we have momentarily restored the gravitational constant $G$ for clarity. This condition is satisfied for astrophysical black holes 
(typically $<10^9 {\rm Gauss}$ \cite{gnedin}). Therefore, the magnetic field can be 
considered as a test field in the background geometry. 
Although the magnetic field is ``weak'' compared with the background, 
it can be quite ``strong'' for charged particles. This can be seen by 
taking the ratio of the Lorentz force  to the gravitational force  acting on a charged particle with charge $q$ and the rest mass $m$   in the Keplerian orbit. For the radius close to 
the Schwarzschild radius, the ratio becomes 
\beqa
\frac{qBM}{m}\sim 10^6 \left(\frac{q}{e}\right)\left(\frac{B}{10^8{\rm Gauss}}\right)\left(\frac{M}{M_{\odot}}\right)\left(\frac{m_p}{m}\right)\,,
\label{ratio}
\eeqa
where $m_p$ is the mass of proton. 
Thus it can be quite large 
for charged particles (protons or electrons) around astrophysical BHs.

\subsection{Black Hole in a Uniform Magnetic Field}

As long as the magnetic field can be treated as a test field, 
we can choose any field configuration we like. 
However, for a Ricci flat spacetime with Killing vectors, 
it is well known that a Killing vector solves the Maxwell equation for
 a 4-vector potential $A^{\mu}$  in the Lorenz gauge: $\nabla_{\mu}A^{\mu}=0$. 
The Kerr spacetime is stationary and 
axisymmetric with Killing vectors $\xi^{\mu}=(\partial/\partial t)^{\mu}$ and $\psi^{\mu}=(\partial/\partial \phi)^{\mu}$. 
Therefore,  $A^{\mu}$ 
is a linear combination of these Killing vectors. 
In particular, as shown by Wald~\cite{wald}, 
for a neutral rotating black hole, the special choice 
\beqa
A^{\mu}=\frac{B}{2}\left(\psi^{\mu}+2a\xi^{\mu}\right)
\label{a:1}
\eeqa
generates an asymptotically uniform  magnetic field of strength $B$.   
However, the second term in Eq.~(\ref{a:1}) is the effect of 
Faraday induction due to the rotation of a BH, which 
generates a difference in the electrostatic potential 
between the event horizon and infinity. Consequently, positively charged particles are accreted towards the horizon. 
For a charged rotating black hole with charge $Q$, 
Eq.~(\ref{a:1}) becomes 
\beqa
A^{\mu}=\frac{B}{2}\left(\psi^{\mu}+2a\xi^{\mu}\right)-
\frac{Q}{2M}\xi^{\mu}\,.
\label{a:2}
\eeqa
Thus, the accretion continues until the potential difference disappears 
 and the black hole will acquire an 
inductive charge of $Q=2a MB$  \cite{wald,aliev}. After 
the accretion is complete,  the 4-vector potential becomes
\beqa
A^{\mu}=\frac{B}{2}\psi^{\mu}\,.
\label{a:3}
\eeqa
Note that as long as the condition Eq.~(\ref{criticalB}) is satisfied, 
the induced charge of the black hole 
is  so small  $Q/M=2aB\leq 2BM\ll 1$ that its effect on the background black hole 
geometry can be neglected.   Hence,  
we shall adopt this choice of 4-vector potential Eq.~(\ref{a:3}) together with the background black hole geometry Eq.~(\ref{kerr}).\footnote{
The motion of charged particles for the choice of the magnetic field Eq.~(\ref{a:1}) was
studied in~\cite{Preti:2010zza}.
}
In Appendix~\ref{appendixa}, we calculate the magnetic flux across a black hole 
for two typical  field configurations Eqs.~(\ref{a:1}) and (\ref{a:3}).

\subsection{Motion of Charged Particles}

The equation of motion for a test particle of  mass $m$ and charge $q$ is given by
\beqa
mu^{\nu}\nabla_{\nu}u^{\mu}=q{F^{\mu}}_{\nu}u^{\nu}.
\label{eom}
\eeqa
Here, $u^{\mu}=\dot x^{\mu}\equiv dx^{\mu}/d\tau$ is the particle 4-velocity 
with $\tau$ being proper time and $u^{\mu}u_{\mu}=-1$.
Also, $F_{\mu\nu}=\nabla_{\mu}A_{\nu}-\nabla_{\nu}A_{\mu}$ is the field strength. 
The equation is derived from the Lagrangian
\beqa
L=\frac12 m g_{\mu\nu}u^{\mu}u^{\nu}+qA_{\mu}u^{\mu}\,,
\eeqa
from which the momentum $p_{\mu}$ conjugate to $x^{\mu}$ is defined by
\beqa
p_{\mu}=mu_{\mu}+qA_{\mu}\,.
\eeqa

For a Kerr black hole immersed in the uniform magnetic field $B(>0)$, 
Killing fields $\xi^{\mu}$ and 
$\psi^{\mu}$ yield a conserved energy per rest mass  $\E$  and an angular momentum per rest mass ${\cal L}$ 
for the motion of a charged particle
\beqa
&&\E=-\frac{1}{m}p_{\mu}\xi^{\mu} =
\left(1-\frac{2Mr}{\Sigma}\right)\dot t+\frac{2aMr\sin^2\theta}{\Sigma}\left(\dot\phi+\frac{b}{2M} \right)
\label{eq:ce}
,\\
&&{\cal L}=\frac{1}{m}p_{\mu}\psi^{\mu}=-\frac{2aMr\sin^2\theta}{\Sigma}\dot t
+\frac{(r^2+a^2)^2-a^2\Delta\sin^2\theta}{\Sigma}\sin^2\theta \left(\dot\phi+\frac{b}{2M} \right)\,,
\label{eq:am}
\eeqa
where we have used Eq.~(\ref{a:3}) and introduced $b=qBM/m$ which is the ratio
in Eq.~(\ref{ratio})~\footnote{Note that our $b$ differs from that in~\cite{fs,zfs} by a factor of $2$, $b_{\rm ZFS}=b/2$.}.
Hence, the azimuthal motion is integrable. 
We note that ${\cal E}$ and ${\cal L}/M$ are dimensionless quantities.
In the presence of a magnetic field,
we should carefully consider the meaning of ${\cal E}$ and ${\cal L}$ 
because
they 
contain the 
magnetic field $b$ in their definition.
Solving Eqs.~(\ref{eq:ce}) and (\ref{eq:am}) in terms of $\dot{t}$ and $\dot{\phi}$, we obtain
\begin{eqnarray}
\dot{t} &=& 
\frac{
\left((r^2+a^2)^2-a^2\Delta\sin^2\theta \right) {\cal E} 
-2 a M r {\cal L} 
}
{\Delta \Sigma},
\label{dottel}
\\
\dot{\phi} &=&
- \frac{b}{2M}
+
\frac{2 a M r {\cal E}  + (-a^2 + \Delta \csc^2\theta)  {\cal L} }
{\Delta \Sigma}.
\label{dotphiel}
\end{eqnarray}
When we discuss the motion of particles on and outside the black hole horizon,
we must impose the forward-in-time condition $\dot{t} \ge 0$ which means that
the value of the time coordinate $t$ increases along the trajectory of the particle.
The radial motion and the polar motion are obtained by solving the equation of motion Eq.~(\ref{eom}),
\beqa
\Sigma \ddot r &=&
-
\frac{a^2 r \sin^2\theta + M(\Sigma-2 r^2)}{\Delta} 
\dot{r}^2
+
a^2
\sin(2\theta) \dot{r}\dot{\theta}
+
\Delta r
\dot{\theta}^2 
-
\frac{b^2 \Delta \sin^2\theta}{4 M^2 \Sigma^2}
(r\Sigma^2 + a^2 M \sin^2\theta(\Sigma-2 r^2))
\nonumber\\
&&
+
\frac{M(\Sigma-2r^2)}{\Delta \Sigma^2}
((a^2 + r^2)\E - a {\cal L})^2
+
\frac{r \sin^2\theta}{\Delta \Sigma^2} (
a(2 M r \E - a {\cal L})
+
{\cal L} \Delta \csc^2\theta)^2,
\label{eomr}
\\
\Sigma \csc(2\theta) \ddot \theta &=&
-\frac{a^2 }{2\Delta}\dot{r}^2
- 
2 r \csc(2\theta)
 \dot{r}\dot{\theta}
+
\frac{a^2}{2} \dot{\theta}^2
- \frac{b^2}{8M^2\Sigma^2}
\left(
\Delta \Sigma^2
+
2Mr(a^2 + r^2)^2
\right)
\nonumber\\
&& + 
\frac{Mr (a \E-{\cal L} \csc^2\theta)^2}{\Sigma^2}
+
\frac{1}{2 \Delta \Sigma^2}(a(2 Mr\E -a{\cal L}) + {\cal L}\Delta \csc^2\theta)^2.
\label{eomtheta}
\eeqa
We solve the above equations 
from a point $x^\mu = x^\mu_{\rm ini}$ 
with an initial velocity $\dot{x}^\mu =u_{\rm ini}^\mu$.
Here we must choose $u_{\rm ini}^\mu$ so that
it satisfies the normalization condition 
$g_{\mu \nu}u_{\rm ini}^\mu u_{\rm ini}^\nu = -1$
and the forward-in-time condition $u^0_{\rm ini} > 0$ if $r_{\rm ini} > r_H$.
Since Eqs.~(\ref{dottel}) - (\ref{eomtheta}) are invariant under the 
transformation $a \to -a, b\to -b, {\cal L} \to -{\cal L}$
and the redefinition of the polar coordinates $\bar{\theta} := \pi - \theta, \bar{\phi} := -\phi$,
we only need to consider $a \ge 0$.
While we need numerics 
to study the
general orbits of charged particle,  
we can solve them analytically  
if we focus on an orbit in the equatorial plane $\theta = \pi/2$, as we will discuss in the next section.

\subsection{ISCO}

\begin{figure}
\includegraphics[height=2.158in]{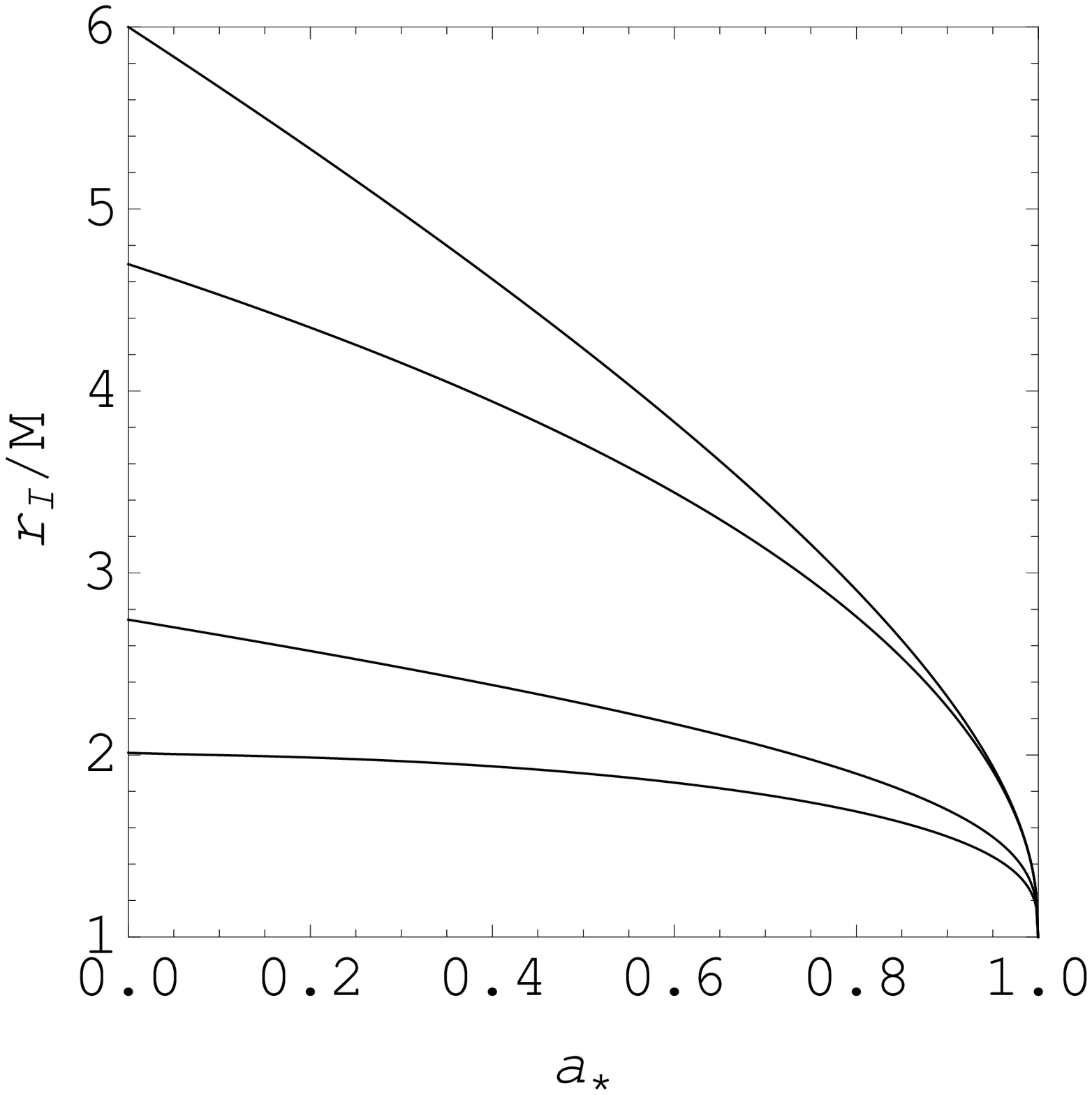}
~
\includegraphics[height=2.158in]{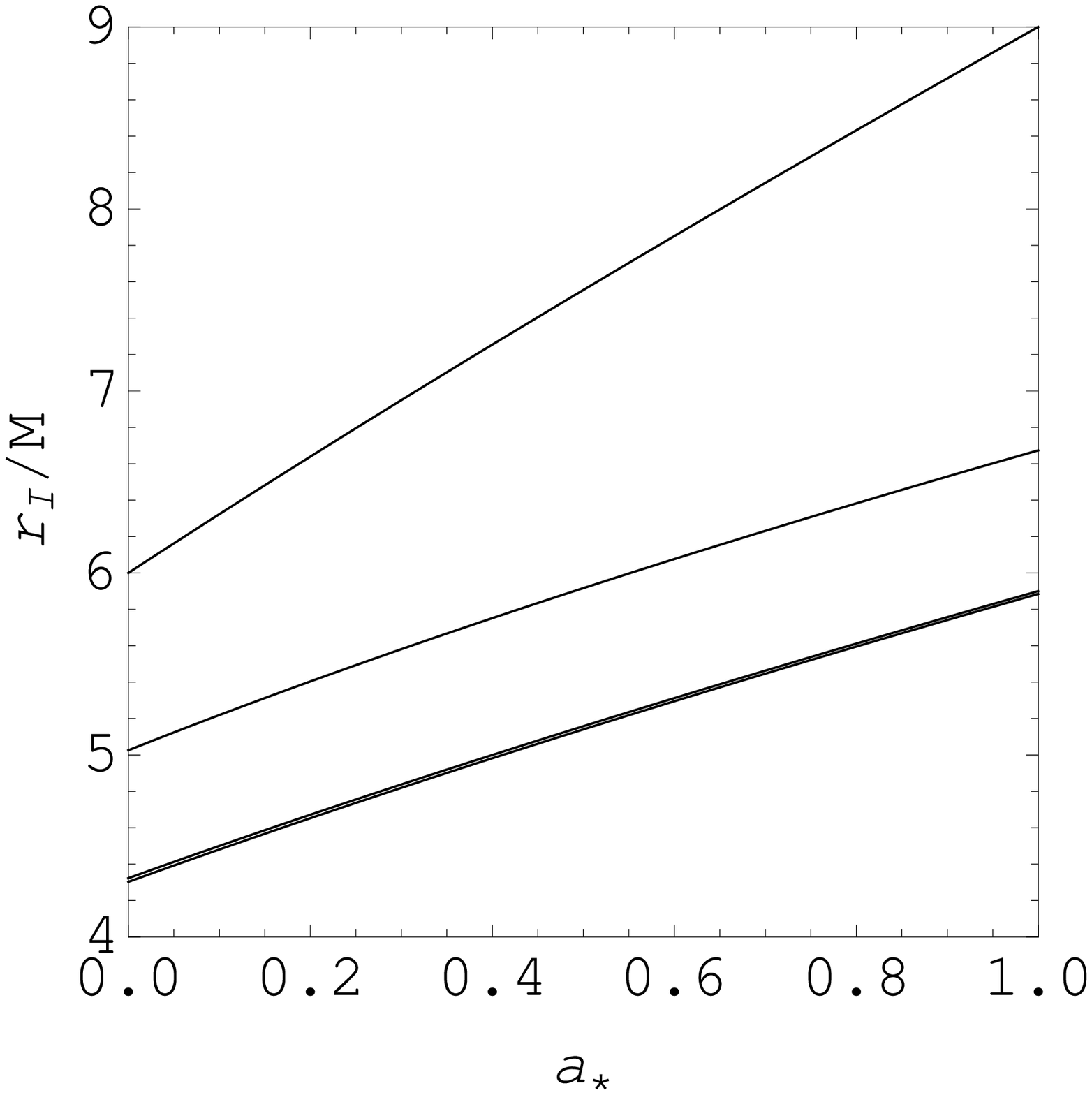}
\\~\\
\includegraphics[height=2.2in]{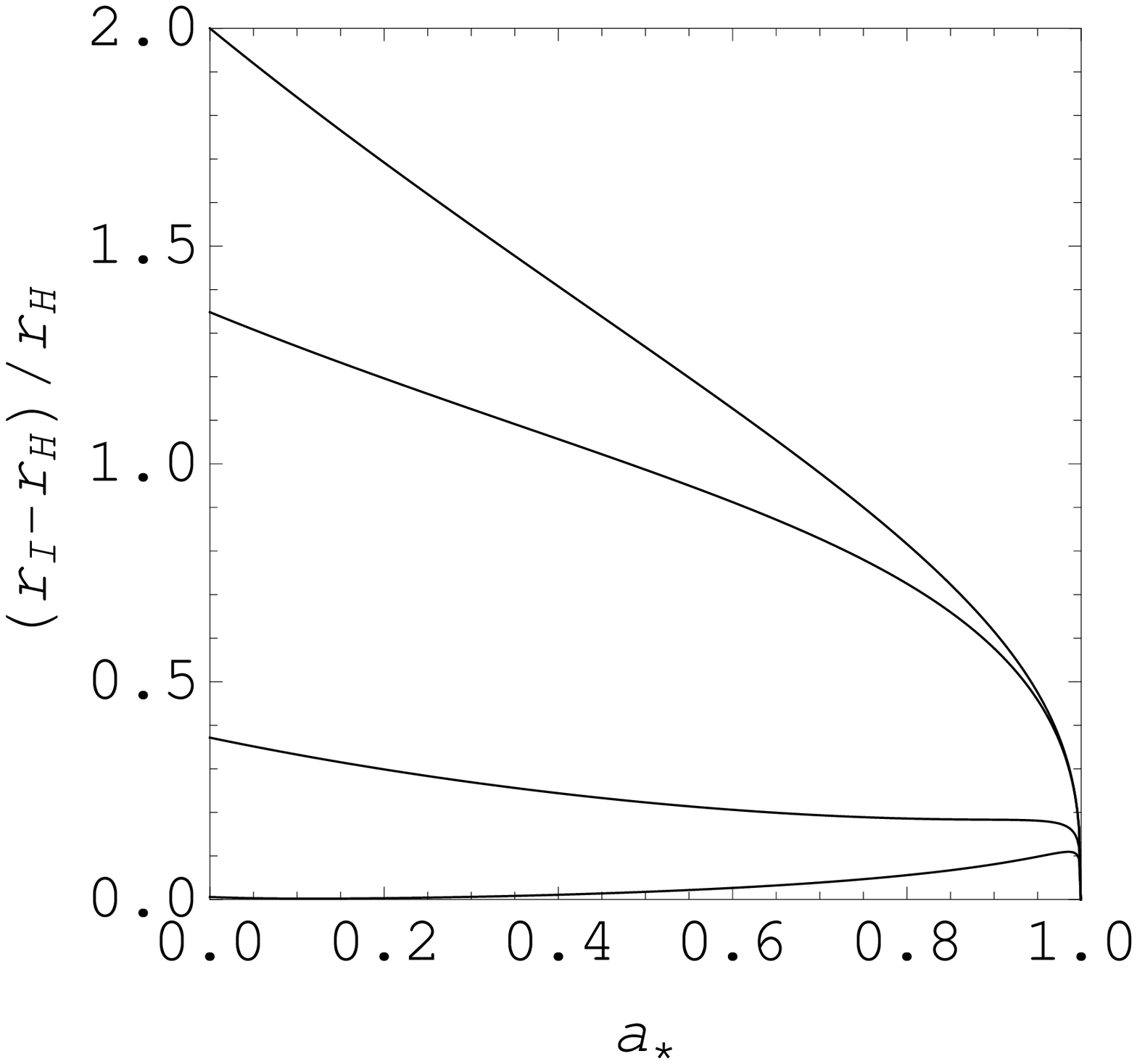}
~
\includegraphics[height=2.2in]{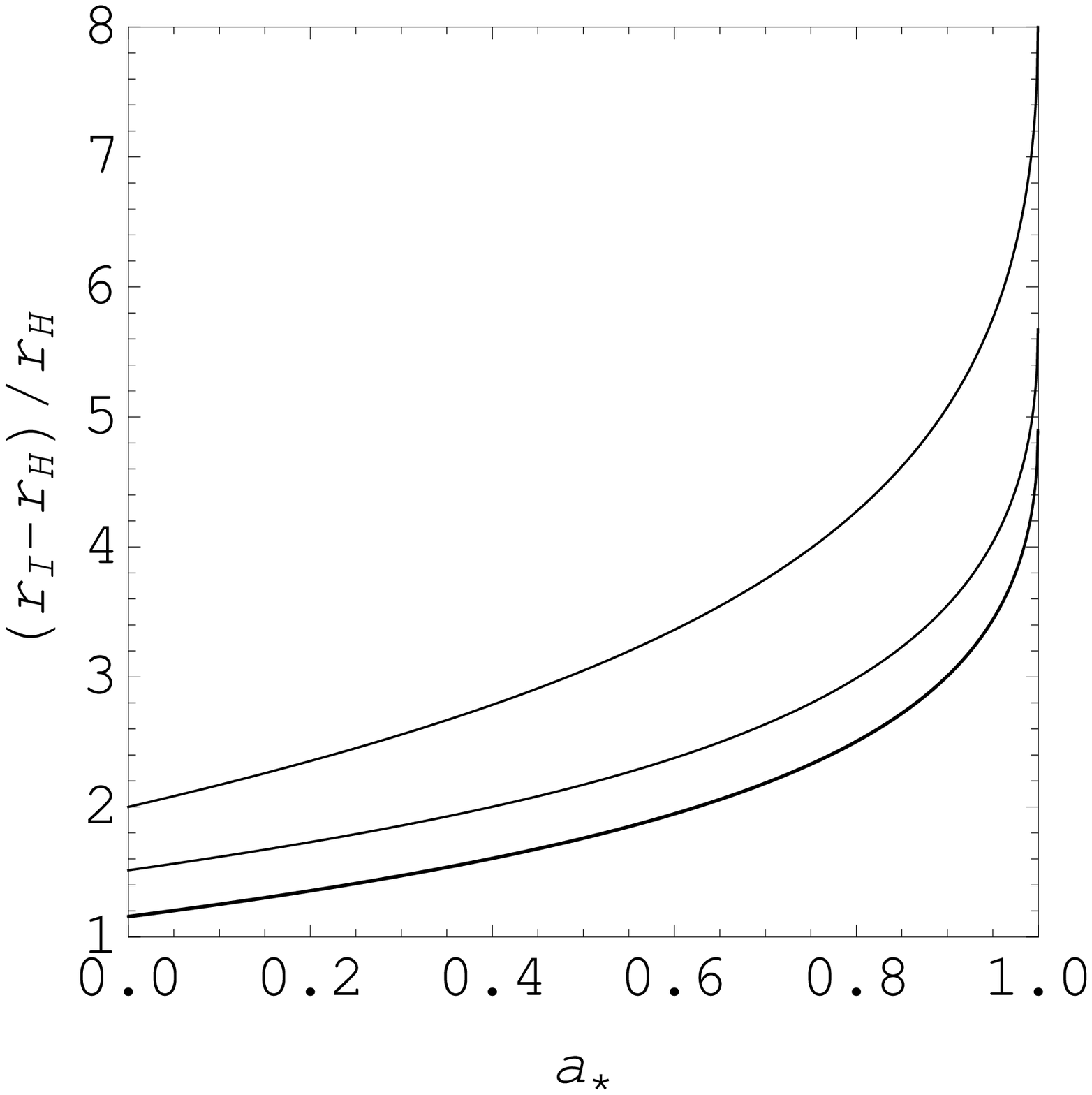}
\caption{\label{fig1}
The ISCO radius $r_I$ as a function of $a_*$ for several $b$. 
The upper two graphs show the dependence of $r_I$ on $a_*$ 
for ${\cal L}>0$ (upper left) and for 
${\cal L}<0$ (upper right), 
and the lower  two graphs represent the difference between $r_I $ and $r_H$
for ${\cal L}>0$ (lower left) and for 
${\cal L}<0$ (lower right).
For all graphs, $b=0,0.1,1,100$ from top to bottom. For ${\cal L}<0$, 
$b=100$ curve almost coincides with $b=10$ curve and is hardly discernible.}
\end{figure}

\begin{figure}
\includegraphics[height=2.0in]{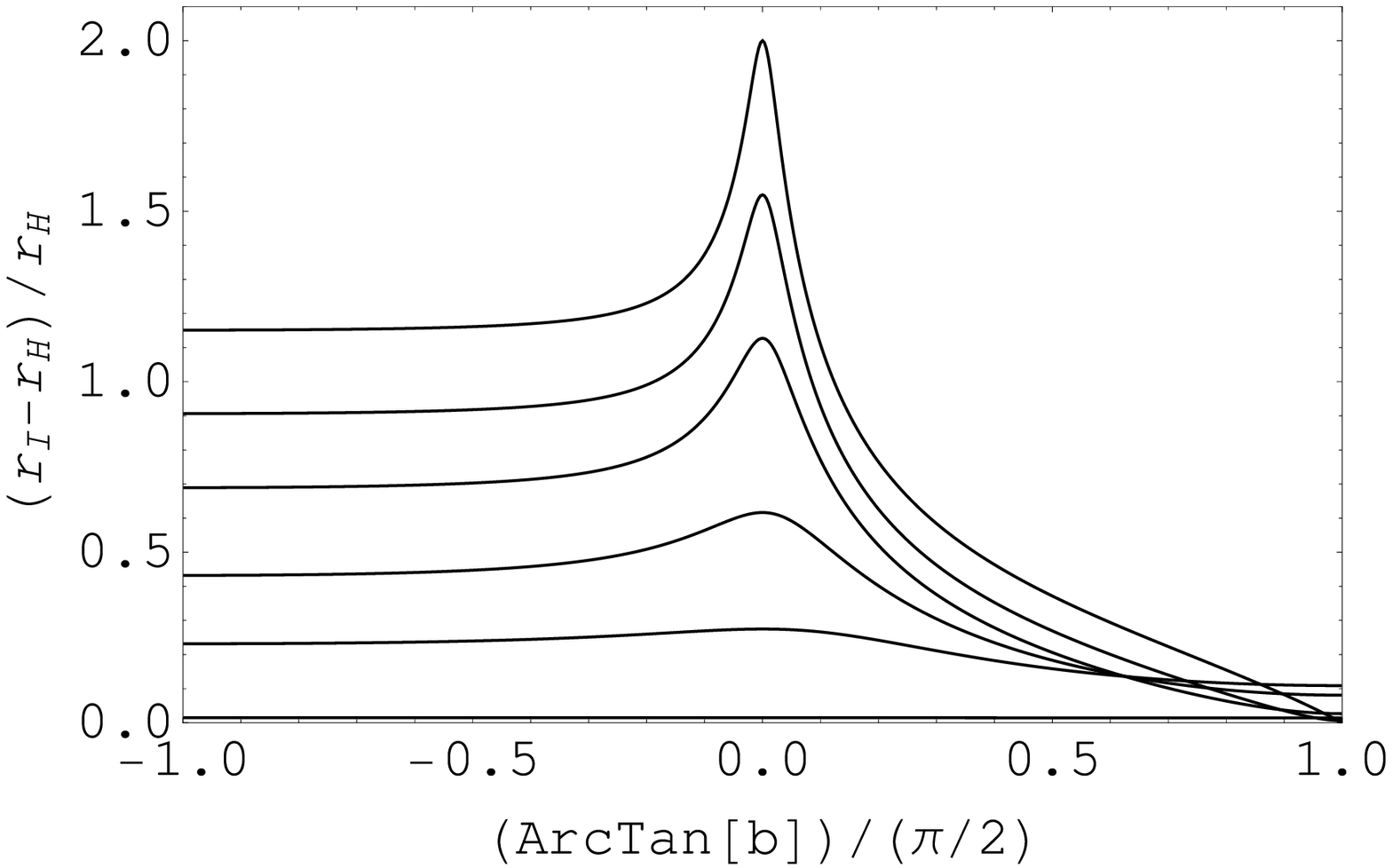}
~
\includegraphics[height=2.0in]{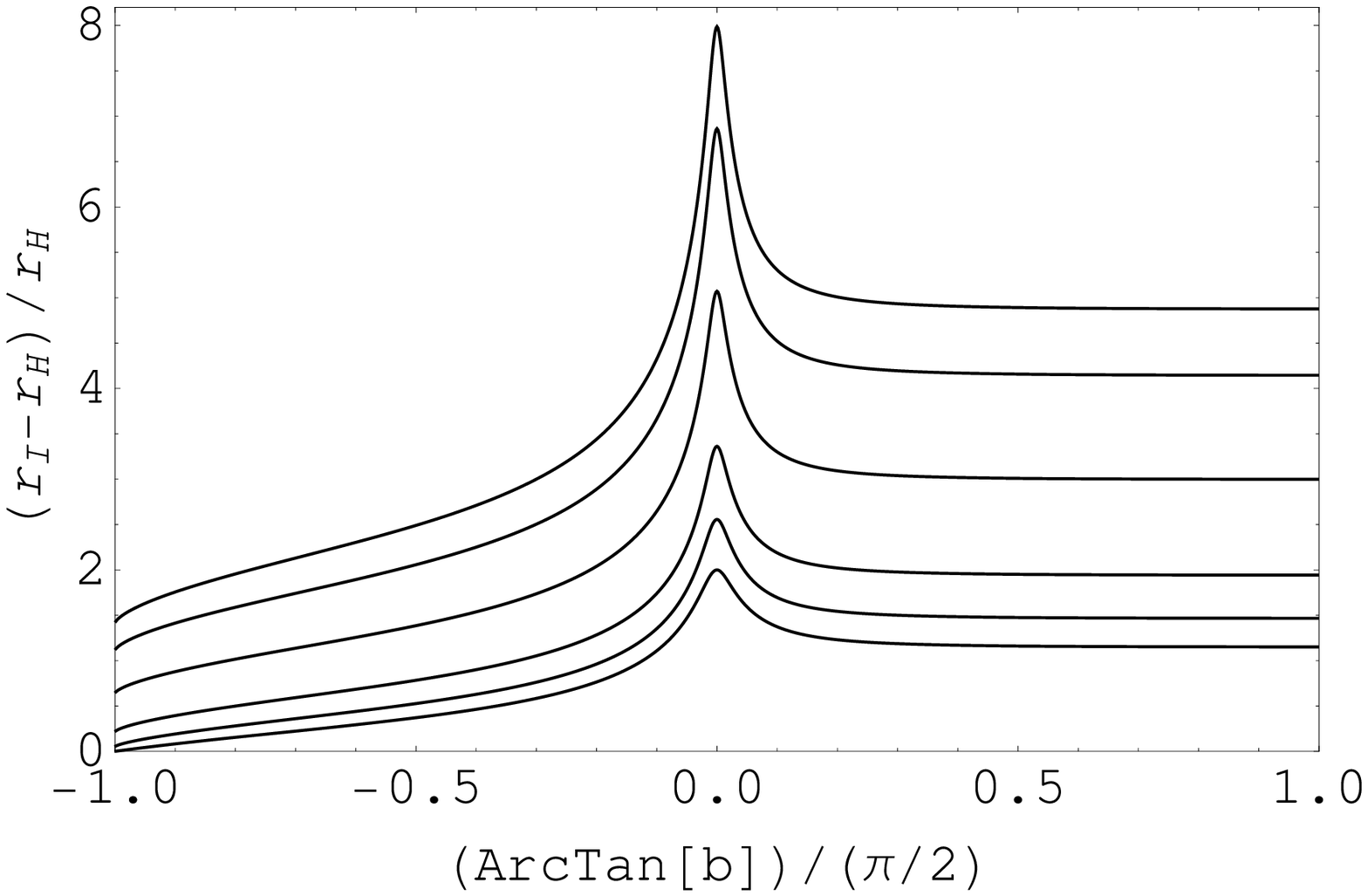}
\caption{\label{fig2}
The ISCO radius $r_I$ as a function of $\arctan b$ for several $a_*$  
for ${\cal L}>0$ (left) and for 
${\cal L}<0$ (right).  For ${\cal L}>0$, $a_*=0,0.3,0.6,0.9,0.99,1$ from top to bottom, 
while from bottom to top for ${\cal L}<0$. }
\end{figure}

Let us consider particle motion in the equatorial plane. 
Then, the equation of motion becomes integrable, and from $u^{\mu}u_{\mu}=-1$ we obtain 
the equation for radial motion~\cite{aliev,igata}
\beqa
r^3 \dot r^2=V(r,\E,{\cal L},b),
\eeqa
where 
\beqa
V(r,\E,{\cal L},b)&=&(r^3+a^2r+2Ma^2)\left(\E^2-\frac{b^2}{4M^2}\Delta \right)
-(r-2M){\cal L}^2
-4Ma\E{\cal L}-r\left(1-\frac{b{\cal L}}{M}\right)\Delta \,.
\eeqa
The maximum of $V$ determines the stable circular orbit and hence $V=\partial V/\partial r=0$ there. 
The innermost of such orbits is called ISCO (the innermost stable circular orbit), 
where relativistic effects 
heavily influences the motion of charged particles.
The ISCO radius is determined by solving the equations 
$V=\partial V/\partial r=\partial^2 V/\partial r^2=0$.  
These were first solved by~\cite{aliev} and the results are
\beqa
&&{\cal L}=-b\left(r-\frac{a^2}{3r}\right)\pm\sqrt{\lambda} \,,
\label{Lisco}\\
&&\E^2=\eta\mp\frac{b}{M}\left(1-\frac{2M}{3r}\right)\sqrt{\lambda} \,,
\label{Eisco}
\eeqa
which are derived from 
$\partial V/\partial r=\partial^2 V/\partial r^2=0$, 
where the upper sign (${\cal L}>0$) refers to  ``prograde'' (or anti-Larmor  according to~\cite{aliev}) motion 
and the lower sign (${\cal L}<0$) refers to  ``retrograde'' (or Larmor) motion.
We note
that the sign of ${\cal L}$ does not necessarily coincide with the sign of $\dot\phi$.  
$\lambda$ and $\eta$ are defined by 
\beqa
&&\lambda=2M\left(r-\frac{a^2}{3r}\right) 
+ 
\frac{b^2}{4M^2}
\left(
r^2\left(5r^2-4Mr+4M^2\right)
+\frac23 a^2(5r^2-6Mr+2M^2)+
a^4\left(1+\frac{4M^2}{9r^2}\right)
\right),\\
&&\eta = 1-\frac{2M}{3r}-
\frac{b^2}{6}\left(
4-5\frac{r^2}{M^2}-\frac{a^2}{M^2}\left(3-\frac{2M}{r}+\frac{4M^2}{3r^2}\right)
\right)\,.
\eeqa
Putting Eqs.~(\ref{Lisco}) and (\ref{Eisco}) into $V=0$ gives the equation for the ISCO radius $r_I$
\beqa
(r^3+a^2r+2Ma^2)\left(\E^2(r)-\frac{b^2}{4M^2}\Delta \right) 
-(r-2M){\cal L}^2(r)-4Ma\E(r){\cal L}(r)
-r\left(1-\frac{b{\cal L}(r)}{M}\right)\Delta &=& 0.
\label{risco}
\eeqa
In general, Eq.~(\ref{risco}) can only be solved numerically. 
We identify the root of Eq.~(\ref{risco}) which is the closest to 
$r_H$ as the ISCO radius $r_I$, although there can be multiple solutions~\cite{aliev}.
We can find the corresponding energy and angular momentum from Eqs.~(\ref{Lisco}) and (\ref{Eisco}). Approximate solutions for limiting values of $a_* \equiv a/M$ and $b$ are given in Appendix~\ref{appendixb}. 
Note that we should exclude the solutions of Eq.~(\ref{risco})
which do not satisfy $\dot{t} > 0$, where $\dot{t}$ is given by Eq.~(\ref{dottel}).
The results of these calculations are given in Figs.~\ref{fig1} and ~\ref{fig2}, where we plot $r_I$ as a function of 
$a_*$ for several $b$ (Fig.~\ref{fig1})
and $r_I$ as a function of $b$ for several $a_*$ 
(Fig.~\ref{fig2}) for both prograde and retrograde motions.  
The left figure in Fig.~\ref{fig2} is the same as \cite{igata}. 
For ISCOs, we find that the sign of ${\cal L}$  coincides with the sign of $\dot\phi$. 

{}From Fig.~\ref{fig1}, we can see that
$r_I$ is uniquely determined by 
 $a_*$ and $b$ in the cases of both prograde 
and retrograde motions.
Focusing only  on the region $b \ge 0$ (or $b < 0$), from Figs.~\ref{fig2} we can see that 
$b$, if it exists, is also uniquely determined by $a_*$ and $r_I$.

\section{Fate of Charged Particles Kicked off from ISCO}
\label{sec3}

We consider the situation where a charged particle is 
initially in the ISCO but acquires a ``kick'' by collisions 
(for example) and then  departs from the equatorial plane. 
The initial velocity  is three-dimensional in general, but in order to 
reduce the space of initial data, we consider as in~\cite{zfs,zahrani} the kick with transverse velocity $v_{\perp}\equiv-r_I\dot\theta$ without changing ${\cal L}$. Even under this restriction, 
the space of the initial data is large enough to find a wide variety of trajectories. 
The problem was studied for a non-rotating black hole in~\cite{zfs}   
and only recently for a rotating black hole in~\cite{zahrani}, 
but the analysis was limited to a fixed value of $b(=0.2)$. 
The slowly rotating case $(a_*=0.5$) was studied by neglecting $O(a_*^2)$  
terms in the equation of motion in~\cite{Hussain:2014cba}.

More concretely,  we numerically solve Eqs.~(\ref{eomr}) and (\ref{eomtheta}) under the initial conditions
\begin{eqnarray}
r_{\rm ini} &=& r_{I},
\quad
\theta_{\rm ini} = \frac{\pi}{2},
\label{initialcondition1}
\\
\dot{r}_{\rm ini} &=& 0, \quad
\dot{\theta}_{\rm ini} = - \frac{v_{\perp}}{r_{I}},
\label{initialcondition2}
\end{eqnarray}
where $v_{\perp} (>0)$ is a constant, and we choose the angular momentum ${\cal L}$
as that of ISCO corresponding to the ISCO radius $r_I$.
The energy ${\cal E}$ is determined from the normalization condition $u^\mu u_\mu = -1$ as 
\begin{eqnarray}
{\cal E} &=& 
\frac{4 a M
   {\cal L} + 
\sqrt{r_I^2 -2 Mr_I + a^2}
\sqrt{4 {\cal L}^2 r_I^2 + (r_I^3 + a^2 (2 M + r_I)) Y }
}{2 \left(r_I^3 + a^2 (2
   M+r_I)\right)},
\label{ethetadot1}
\end{eqnarray}
with
\begin{eqnarray}
Y &=& 
4  \dot{\theta}_{\rm ini}^2 r_I^3   
+
(b/M)^2 \left(r_I^3 + a^2 (2
   M+r_I)\right)
+
4 r_I(1 - b{\cal L}/M).
\label{ethetadot2}
\end{eqnarray}

For a given  $a_*, b$ and $v_{\perp}$, we solve Eqs.~(\ref{eomr}) and (\ref{eomtheta}) under the above initial conditions. 
We assume $b>0$ in the following. 
We checked the accuracy of our numerical calculations by verifying constancy of $\E$.

\begin{figure}
\includegraphics[height=3.5in]{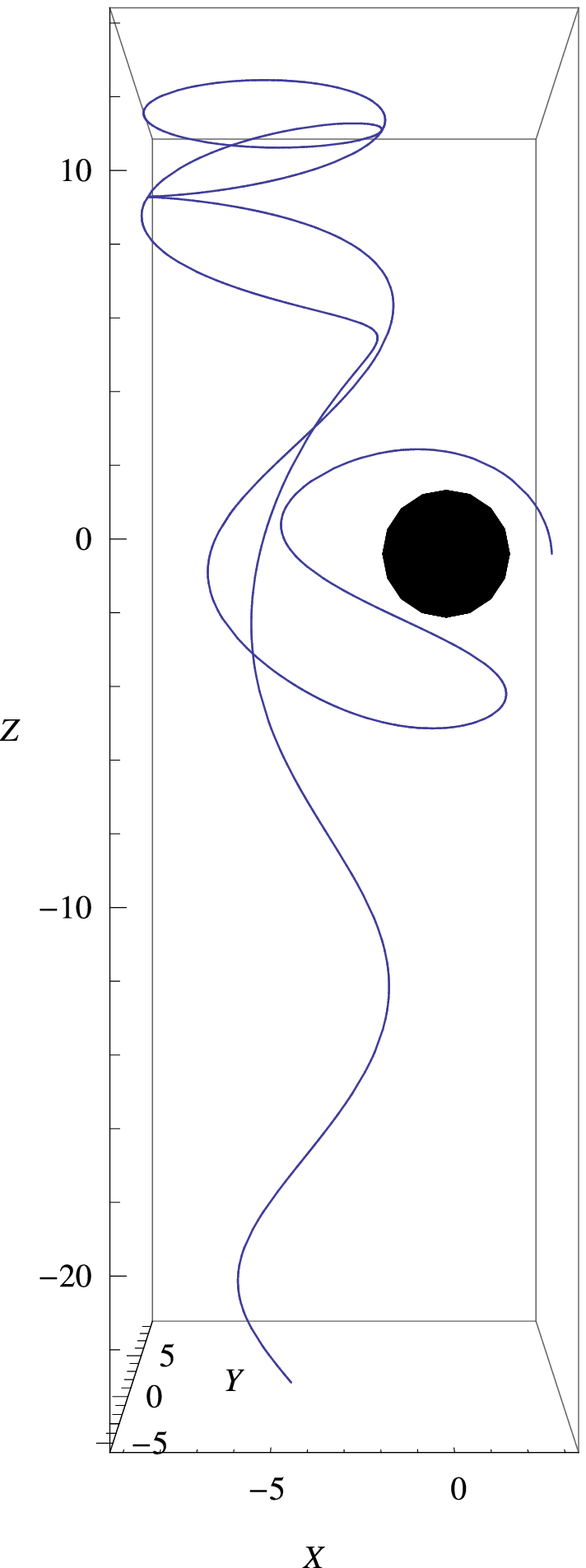}
~~~~~
\includegraphics[height=3.5in]{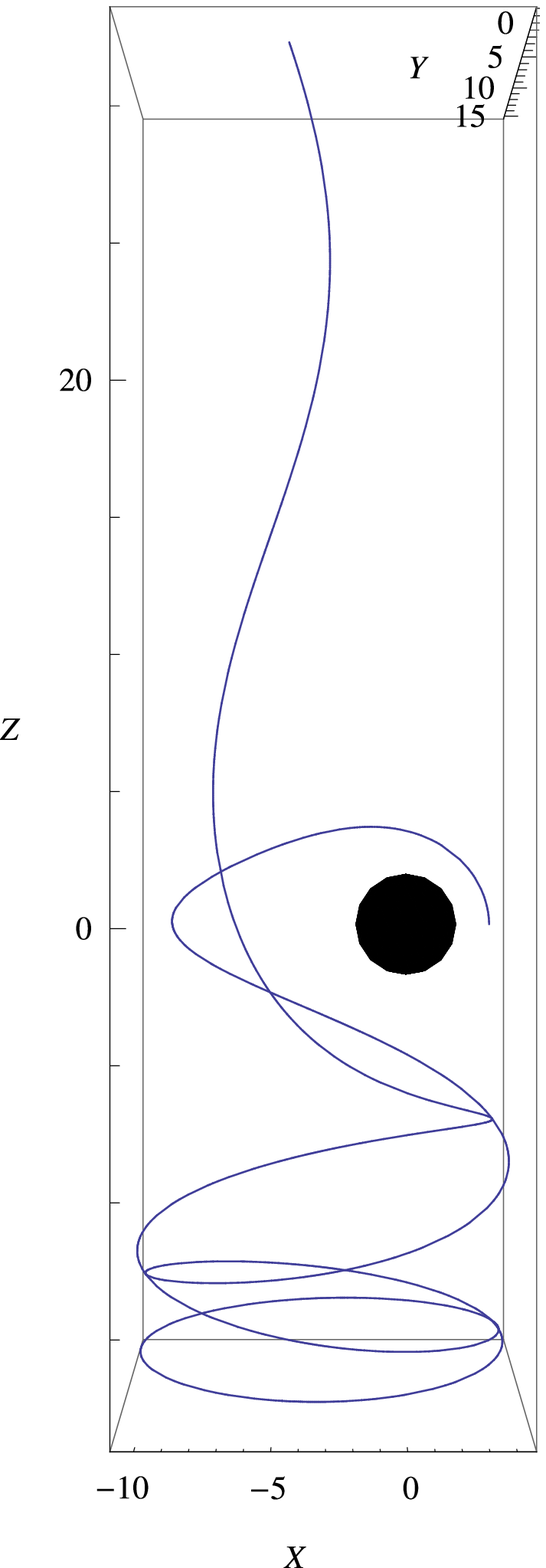}
~~~~~
\includegraphics[height=1.7in]{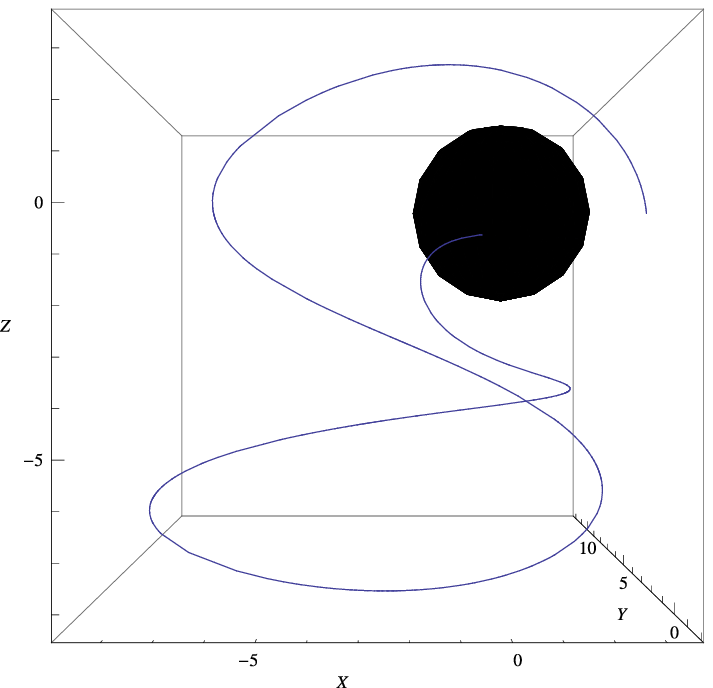}
~~~~~
\includegraphics[height=0.9in]{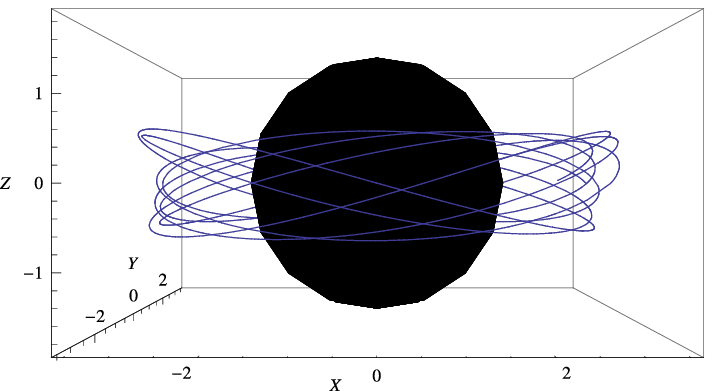}
\caption{\label{fig3DLp}
The typical trajectories of the charged particle kicked off from ISCO for prograde motion $({\cal L}>0)$.
We set the parameters as $a_* = 0.5, b=0.24$, and the corresponding ISCO radius is $r_I/M = 3.1081$.
The figures are that of 
$z \to -\infty$ orbit for ${\cal E} = 1.24058$ (left),
$z \to \infty$ for ${\cal E} = 1.90728$ (middle left),
capture orbit for ${\cal E} = 1.56367$ (middle right)
and 
bound orbit for ${\cal E} = 0.761303$ (right), respectively.
}
\end{figure}

\begin{figure}
\includegraphics[width=0.35\linewidth,clip]{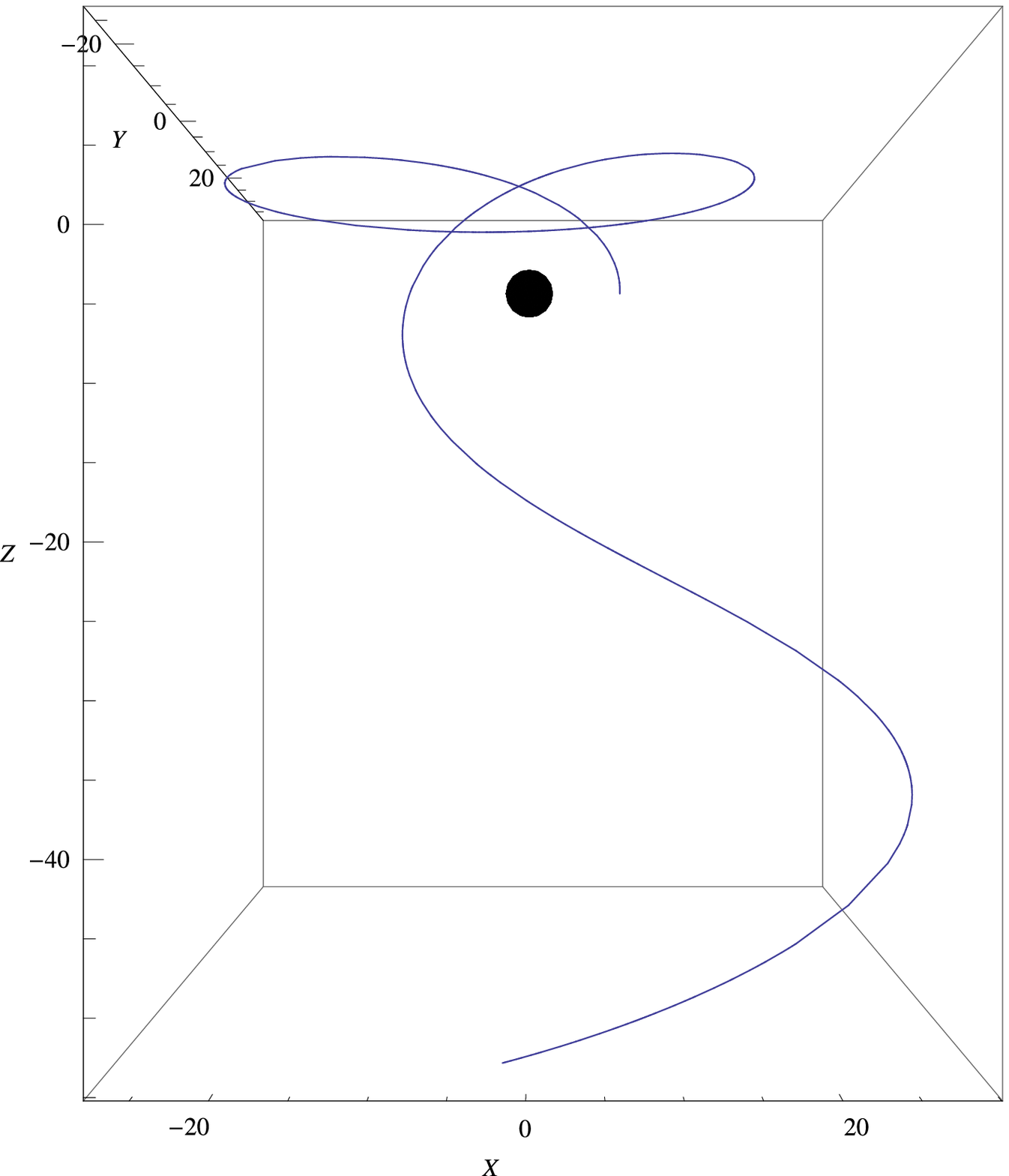}
~~~~~
\includegraphics[width=0.35\linewidth,clip]{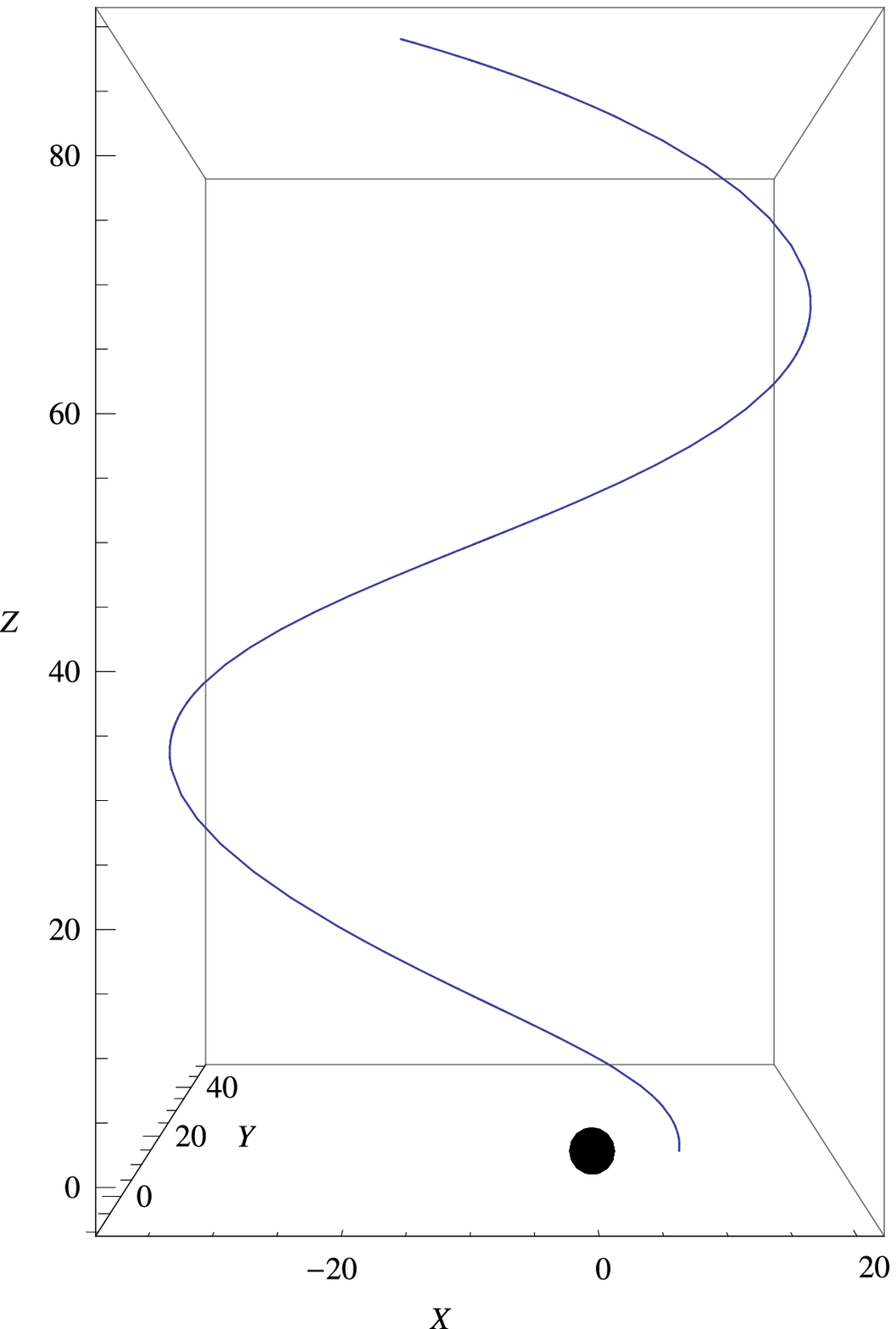}
\\~\\
\includegraphics[width=0.4\linewidth,clip]{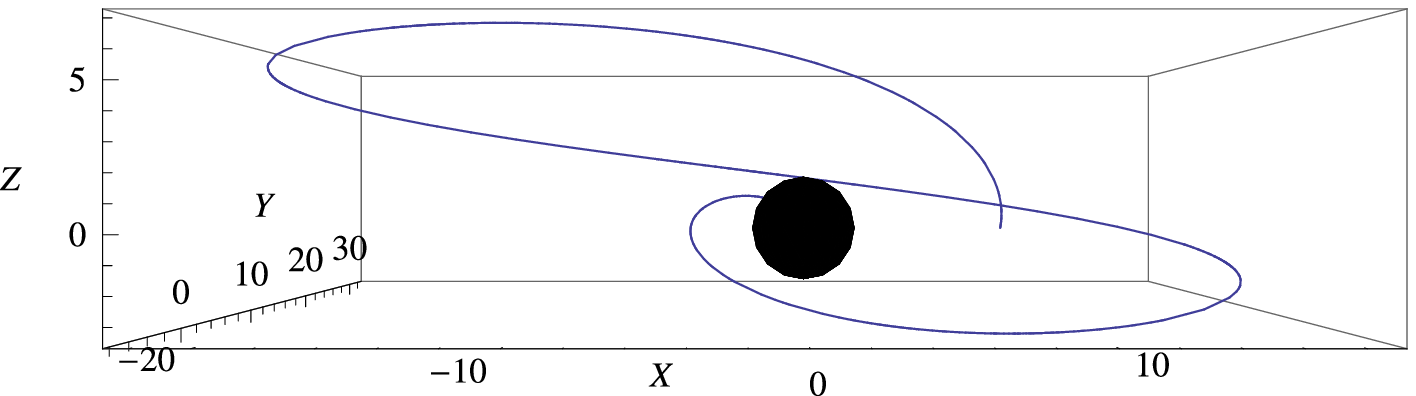}
~~~~~
\includegraphics[width=0.4\linewidth,clip]{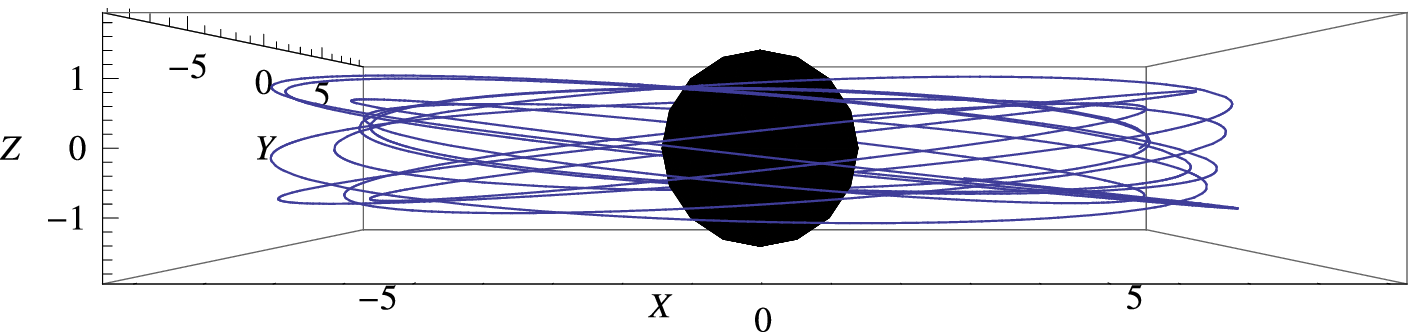}
\caption{\label{fig3DLn}
The typical trajectories of the charged particle kicked off from ISCO for retrograde motion $({\cal L}<0)$.
We set the parameters as $a_* = 0.5, b=0.02$, and the corresponding ISCO radius is $r_I/M = 7.2223$.
The figures are that of 
$z \to -\infty$ orbit for ${\cal E} = 1.08614$ (upper left),
$z \to \infty$ for ${\cal E} = 1.17061$ (upper right),
capture orbit for ${\cal E} = 1.56367$ (lower left)
and 
bound orbit for ${\cal E} = 1.00021$ (lower right), respectively.
}
\end{figure}

We find that there are four different final states 
for the particle:  
capture by the black hole,  escape to $z \to \pm \infty$, 
and bound motion. 
In our calculations, the maximum integration time was chosen to be
$\tau = \tau_{\rm max} =2 M \times 10^{5}$. 
We consider the particle to have ``escaped'' if $|z|>10^3 M$, ``captured'' when $r$ reaches $r_H$, or otherwise in a ``bound orbit''. 
Typically, the error in the energy is less than $10^{-6}$, but sometimes grows to 
$10^{-2}$ when
the integration time is very long, which is the case with escape to $|z|\rightarrow \infty$ 
(the increase of the error 
for the escape orbit was also discussed in~\cite{zfs}). 
The typical trajectories of the charged particle are 
depicted in Figs.~\ref{fig3DLp} (for ${\cal L}>0$) and \ref{fig3DLn} (for ${\cal L}<0$).


\begin{figure}
\includegraphics[height=2.15in]{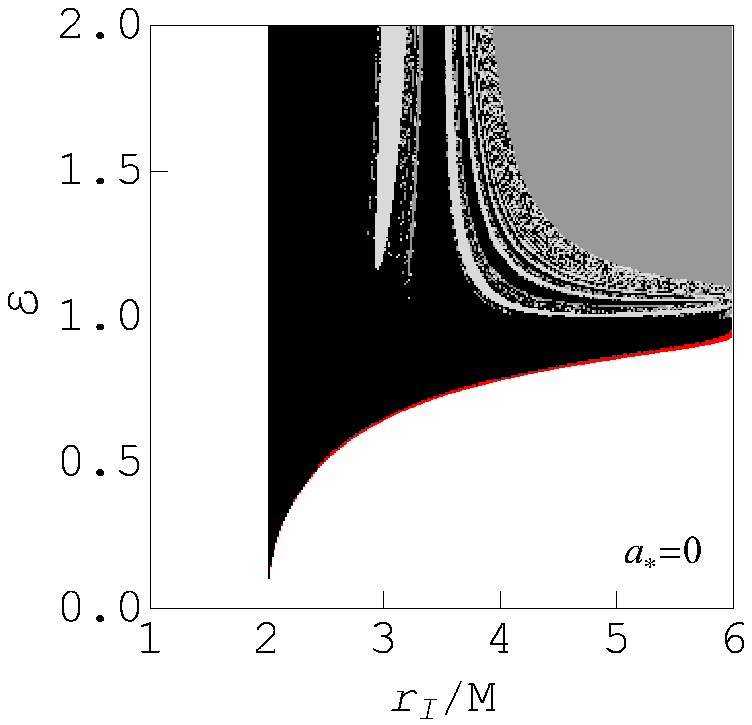}
~
\includegraphics[height=2.15in]{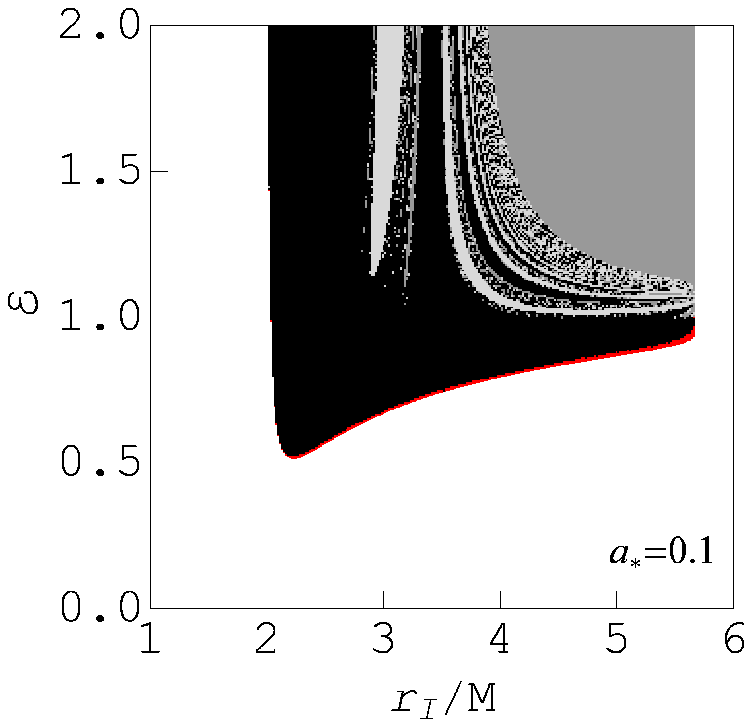}
~
\includegraphics[height=2.15in]{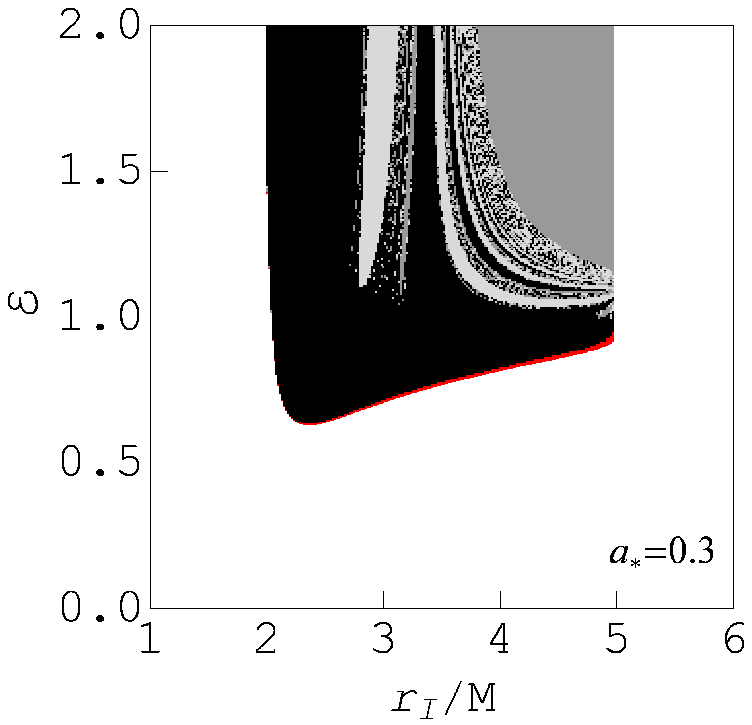}
\\~~\\
\includegraphics[height=2.15in]{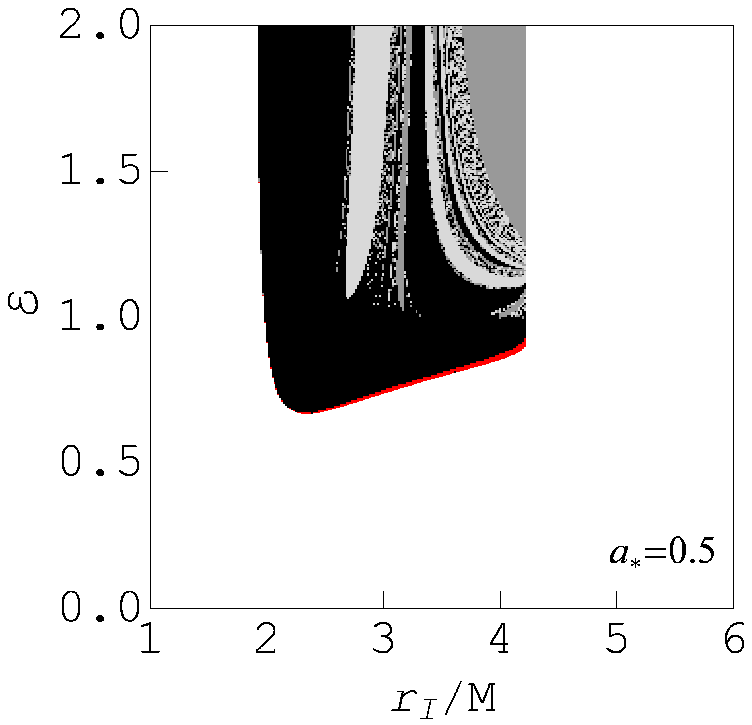}
~
\includegraphics[height=2.15in]{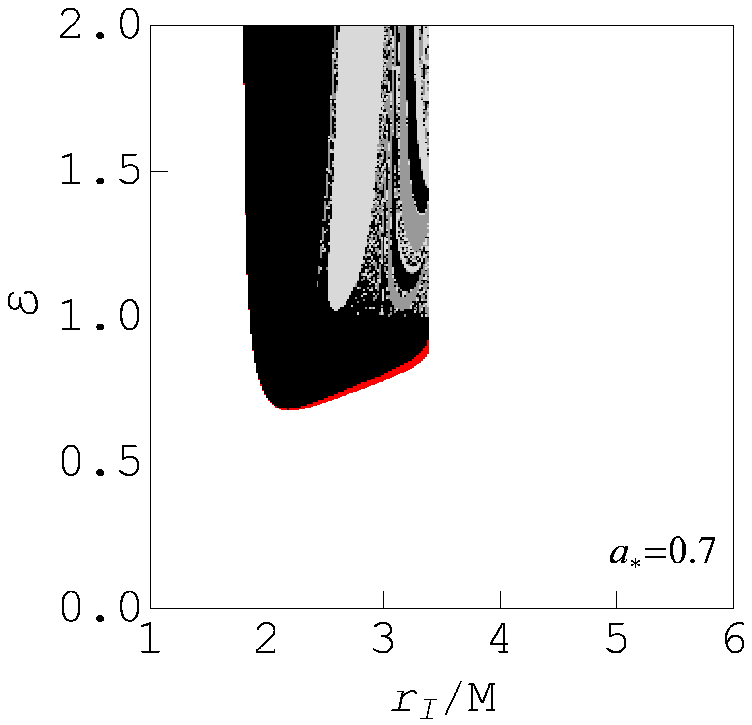}
~
\includegraphics[height=2.15in]{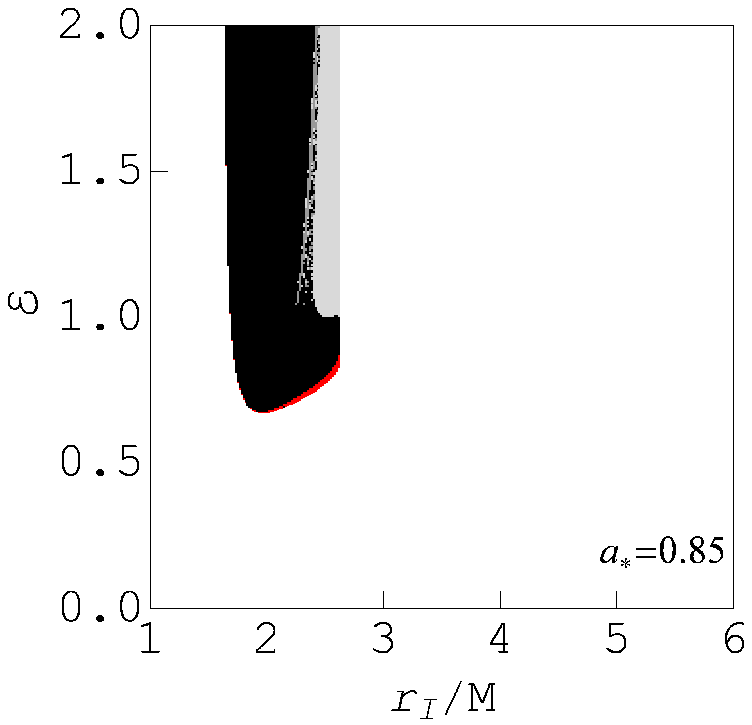}
\\~~\\
\includegraphics[height=2.15in]{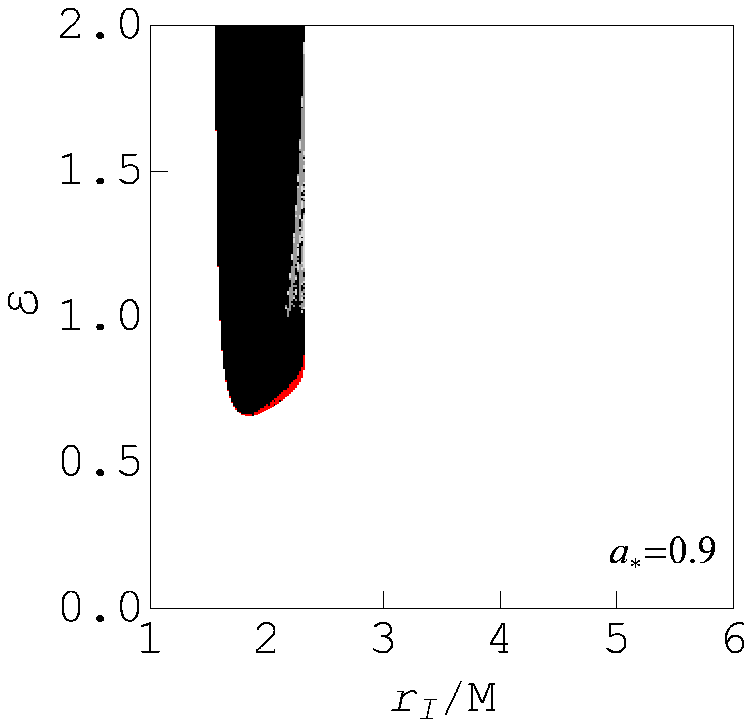}
~
\includegraphics[height=2.15in]{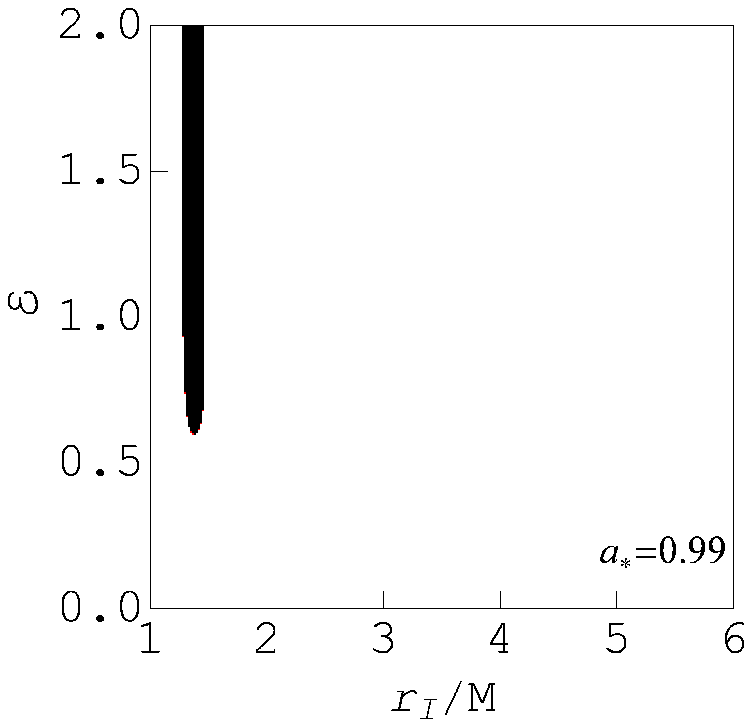}
\caption{\label{fig4}
Fate of charged particles
kicked off from the prograde ISCO (${\cal L}>0$) 
for $a_* = 0, 0.1, 0.3, 0.5, 0.7, 0.85, 0.9, 0.99$ from left to right and down.
The dots represent capture (black), 
escape to $z\to \infty$ (gray),
escape to $z\to -\infty$ (light gray) 
and bound motion (red), respectively.
No allowed motion in ISCO in the white area.}
\end{figure}

\begin{figure}
\includegraphics[height=2.1in]{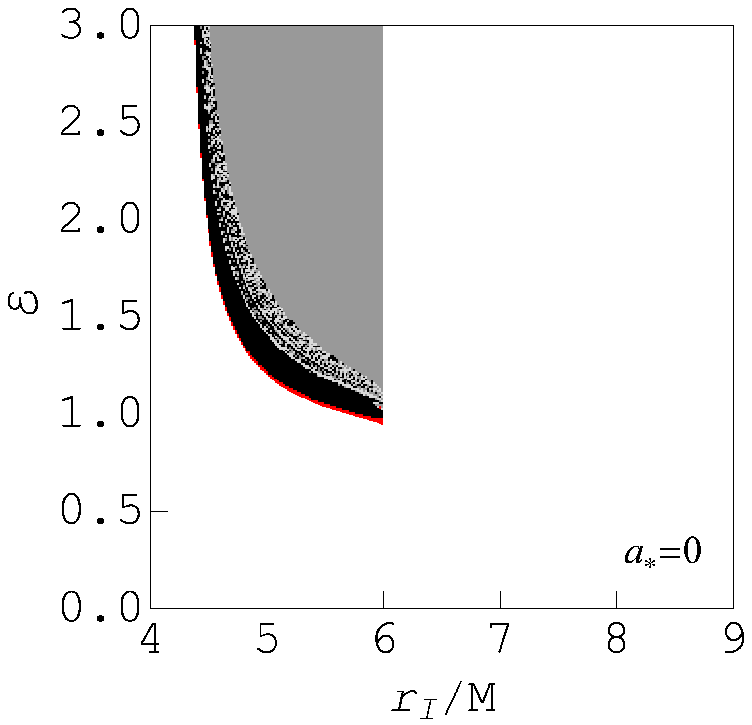}
~~
\includegraphics[height=2.1in]{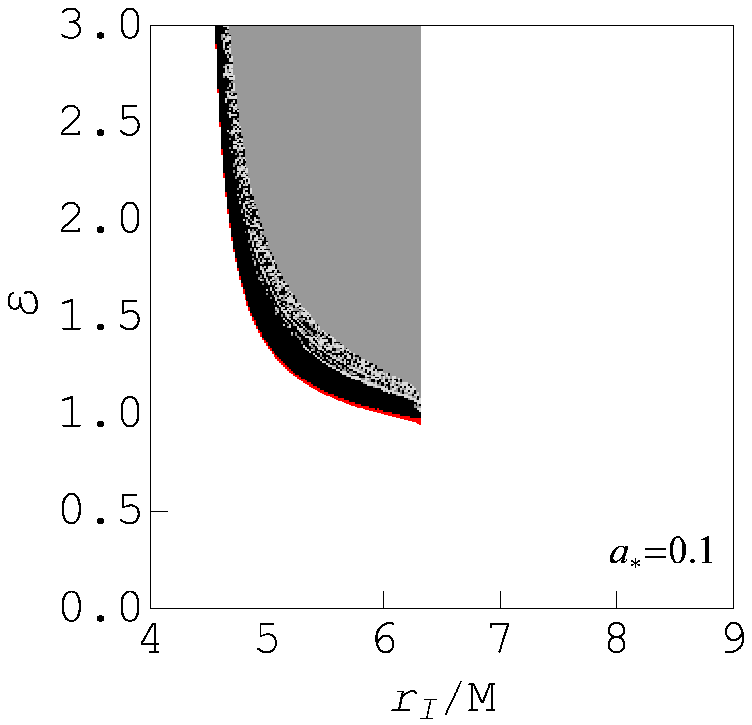}
~~
\includegraphics[height=2.1in]{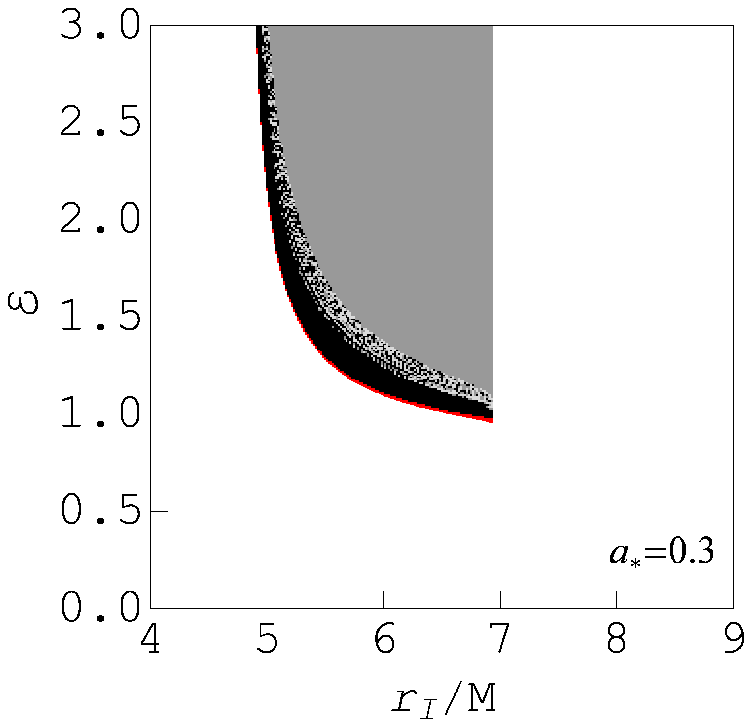}
\\~~\\
\includegraphics[height=2.1in]{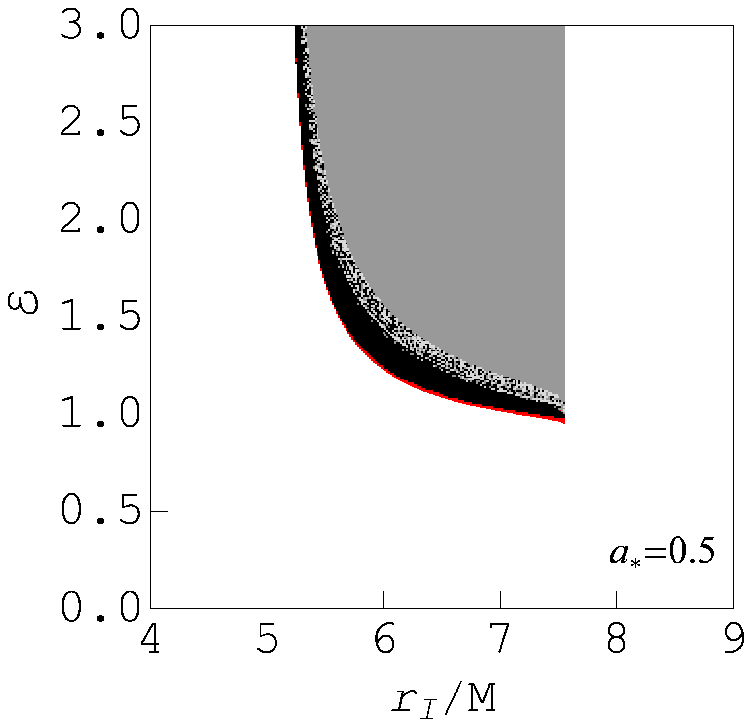}
~~
\includegraphics[height=2.1in]{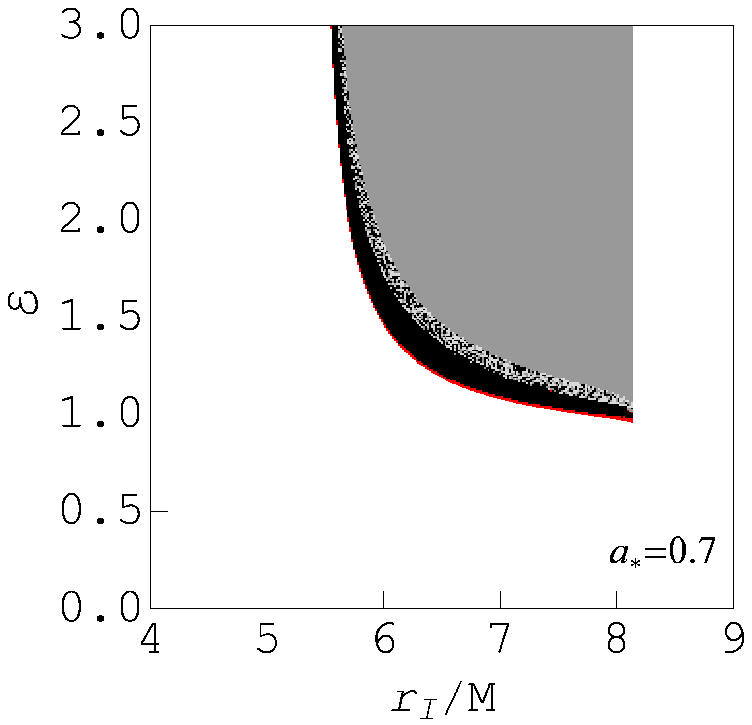}
~~
\includegraphics[height=2.1in]{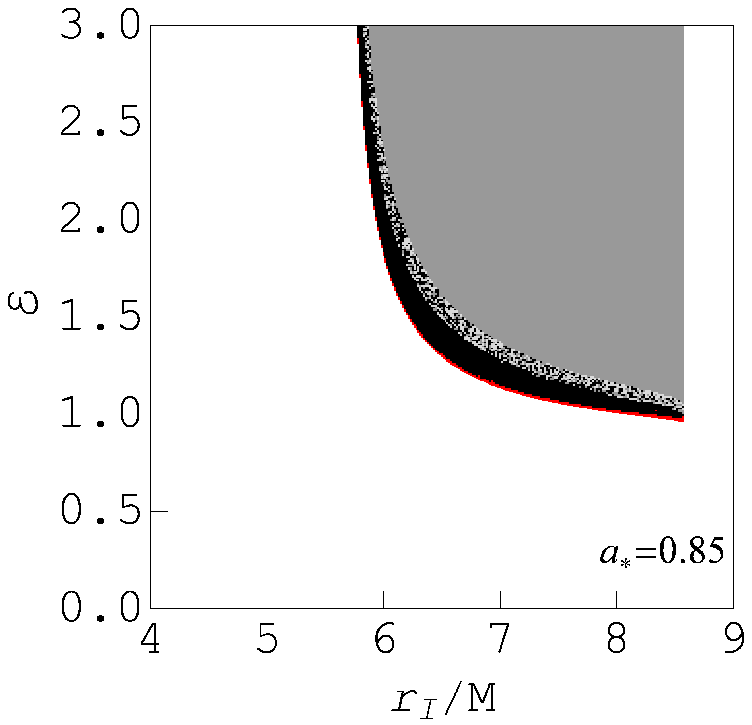}
\\~~\\
\includegraphics[height=2.1in]{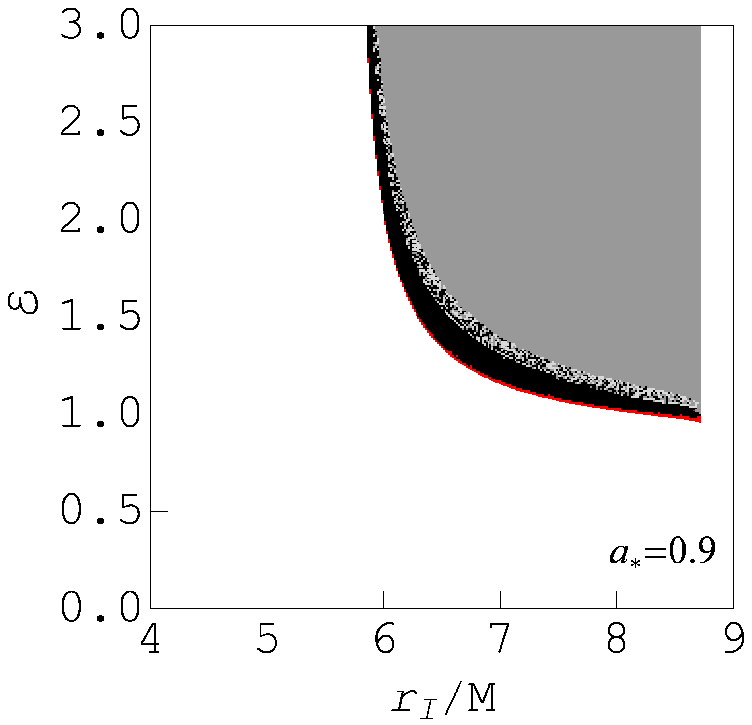}
~~
\includegraphics[height=2.1in]{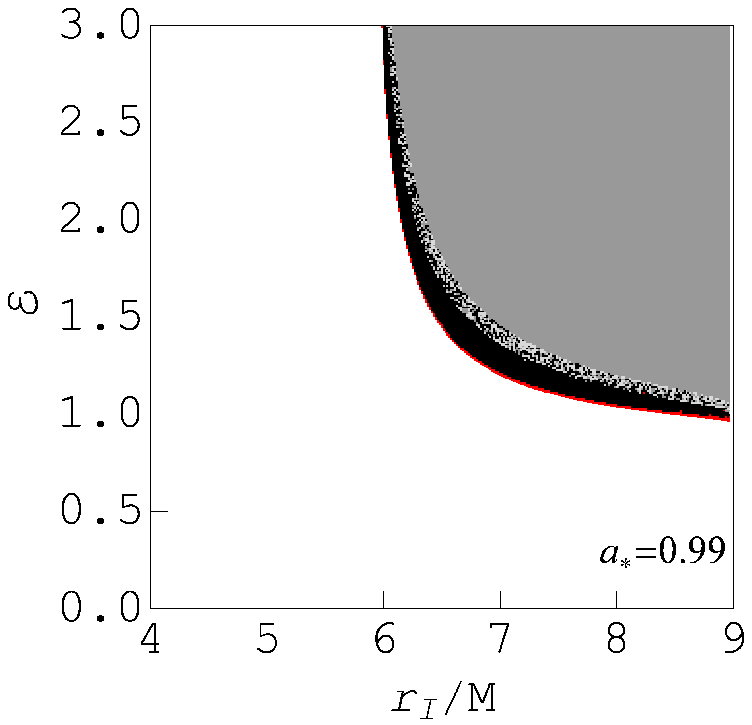}
\caption{\label{fig5}
The same as Fig.~\ref{fig4}, but for the retrograde ISCO (${\cal L}<0$). }
\end{figure}

Fig.~\ref{fig4} shows the basin of attraction 
for ${\cal L}>0$ for several $a_*$. Fig.~\ref{fig5} is for ${\cal L}<0$. 
The horizontal axis denotes the ISCO radius $r_I$ normalized by $M$ for $b(> 0)$ and
the vertical axis denotes the energy ${\cal E}$ which is determined from $v_{\perp} = - r_H \dot{\theta}_{\rm ini}$ 
using Eqs.~(\ref{ethetadot1}) and (\ref{ethetadot2}). 
We note that $r_I$ is a function of $b$ for a fixed $a_*$ 
as shown in Fig.~\ref{fig2}.
The resolution of the plots in these figures is $300\times 300$. 
The color of each dot in these figures determines the fate of 
the particle motion: black for capture, gray for escape to $z\rightarrow \infty$, light gray for escape to  
$z\rightarrow -\infty$, and red
for bound motion. 
The white areas correspond to regions forbidden for ISCO orbits.
The top left graphs in Figs.~\ref{fig4}-\ref{fig5} are for $a_*=0$  
and agree with those in~\cite{zfs}.


{}From  Fig.~\ref{fig4}, 
we see that the allowed region becomes smaller as $a_*$ increases.
This is because  $r_I$ decreases 
as $a_*$ increases for a fixed $b$.
In particular,
$r_I|_{b =0}$ decreases at an almost constant rate
but $r_I|_{b = \infty}$ almost coincides with $r_{H}$, 
irrespective of $a_*$ (see Fig.~\ref{fig1}).
We also find that for the same ${\cal E}$ and $r_I$, 
the fates of the charged particles are almost the same. 
The allowed region is  gradually ``eaten''
with increasing $a_*$. 
In the $a_*=0$ figure (top left), the left area (black) 
corresponds to the region near the black hole horizon, so 
the orbits are almost all captured.
In the right area (gray), $r_I$ is much larger $r_H$ and 
the effect of the gravity is relatively weak, so 
the particle can escape to $z \to \infty$ for 
sufficiently large $v_{\perp}(>0)$
as in the case of Minkowski spacetime. 
The intermediate region looks rather complicated, 
but it was shown in~\cite{zfs,zahrani} 
that the basin of attraction is a fractal.
The effect of increasing $a_*$ is to cut the allowed region 
for $a_*=0$ from the right.

Similarly, we can understand the features of Fig.~\ref{fig5}. 
This time, the allowed region gets larger as $a_*$ increases. 
This is because for retrograde ISCOs, $r_I$ increases 
as $a_*$ increases for a fixed $b$
in contrast to the case of prograde ISCOs.
In this case, both $r_I|_{b =0}$ and $r_I|_{b = \infty}$ 
increase and the difference between them also increases 
as $a_*$ increases (see Fig~\ref{fig1}).
Moreover, since $r_I$ of the retrograde motion is larger than 
that of the prograde motion, 
the effect of the gravity is rather weak for any $a_*$.
Hence, we expect that the particle can escape 
to $z \to \infty$ for  large $v_{\perp}$.
Thus, the allowed region becomes larger and shifts
toward right  as $a_*$ increases.

In Figs.~\ref{fig4}-\ref{fig5}, 
we plot several red dots between black and white regions.
These dots correspond to the bound motion; 
if the energy is close to that of ISCO, the particle neither escapes to $z \to \pm \infty$ 
nor is captured by the black hole 
until at least $\tau = \tau_{\rm max} =2 M \times 10^{5}$
(these orbits are 
also observed in~\cite{zahrani}). 
These bound orbits are located around ISCO orbits as shown  
in the right of Fig.~\ref{fig3DLp}.
For some of these  red dots, 
we checked that the motion remains bound  
even if we extend the maximum integration time to 
$10\times \tau_{\max} = 2M \times 10^6$. The energy error 
was less than $1.3\times 10^{-3}$. 
We note
that 
there might exist {\it quasi}-bound orbits around 
ISCO which survive for a long time, which
may have implications for the high energy particle collision scenario proposed in~\cite{igata}. 
We leave the detailed analysis for future work.

\section{Summary} 
\label{secsum}

We have studied the motion of charged particles around a weakly magnetized 
rotating black hole. First, we have studied in detail the effects of black hole spin 
and an external magnetic field on the ISCOs of charged particles. We found that 
the radius of the ISCO decreases as the magnetic field increases. 
Next, we have studied the motion of a charged particle kicked out from the ISCO. 
We found that trajectories of the particle are full of variety.  
However, the asymptotic behavior is classified into four types: capture by the black hole, 
escape to $z\rightarrow \pm \infty$, and the bound motion.  
We found that the final fate depends on the energy of the particle and mainly on 
the radius of ISCO.  
The energy and the radius depend on the initial velocity, the black hole spin, and the magnetic field. 
According to 
our numerics, 
particles in bound motion stay
in the vicinity of the equatorial plane.

It would be interesting to study the possible existence and stability of bound orbits near 
the equatorial plane  which may widen the region where high energy 
particle collisions take place~\cite{igata}. 
It would also be important to study particle motion in other field configurations 
and examine the robustness of our results.

\section*{ACKNOWLEDGEMENTS}

We would like to thank B.~Way and the referee 
for their careful reading of the manuscript and useful comments. 
We also thank 
the Yukawa Institute for Theoretical Physics at Kyoto University 
where this work was initiated during the YITP workshop YITP-X-13-03 on 
``APC-YITP collaboration: mini-workshop on gravitation and cosmology''.
This work is supported by the grant for research abroad 
from JSPS (MK) and by the Grant-in-Aid for Scientific Research
from JSPS (Nos.\,24540287 (TC)) and in part
by Nihon University (RS, TC).

\appendix

\section{Magnetic Fluxes Across Black holes}
\label{appendixa}

In this appendix, we calculate the flux of an asymptotically uniform  magnetic  field 
across one half of the horizon of a rotating black hole. 
The flux of a magnetic field threading the upper half of the horizon is given by~\cite{klk}
\beqa
\Phi=\int^{2\pi}_{0}d\phi\int^{\pi/2}_{0}d\theta F_{\theta\phi}|_{r=r_H}\,.
\eeqa
For a charge neutral black hole, the 4-vector potential is  Eq. (\ref{a:1}) and the 
flux is well-known~\cite{klk} and is given by
\beqa
\Phi=\pi Br_H^2\left(1-\frac{a^4}{r_H^4}\right)=4\pi BM\sqrt{M^2-a^2}\,.
\eeqa
The flux decreases as $a$ increases, which is sometimes called a 
``Meissner-like'' effect. 
However, for a black hole with vanishing electrostatic potential, 
the 4-vector potential is  Eq.~(\ref{a:3}) and this time the flux is given by
\beqa
\Phi=4\pi BM^2\,,
\eeqa
which is {\it independent} of $a$. Hence, the presence of a 
``Meissner-like'' effect depends on 
the choice of the field configuration and does not occur 
in general.

\section{Approximate solutions for ISCO}
\label{appendixb}

Although Eq.~(\ref{risco}) can only be solved numerically 
for general $a_*$ and $b$, it can be solved analytically for 
limiting values of $a_*$ and $b$. However, the previous analyses were  limited to 
the ISCO of a maximally rotating black hole 
($a_*=1$)~\cite{aliev} or to the prograde orbit for a nearly maximally rotating 
black hole~\cite{igata}. We extend these analyses 
to include retrograde motion.

\subsection{ISCO for prograde motion}

For a maximally rotating black hole 
($a_*=1$), 
the radius of the ISCO for a prograde motion  (${\cal  L}>0$) is given by~\cite{aliev}
\beqa
r_I^{a_*=1}/M=1\,,
\label{isco:l>0}
\eeqa
independently of $b$. 
For a nearly maximally rotating black hole ($a_*\simeq 1$),  
the correction to  Eq. (\ref{isco:l>0}) is \cite{igata}
\beqa
r_I/M-r_I^{a_*=1}/M &=& 2^{2/3}(1-a_*)^{1/3} 
+
\frac{7+b^2\left(5-8 b^2-6b \sqrt{3+4 b^2}\right) }{2^{5/3}\left(1+b^2\right)^2}(1-a_*)^{2/3}+O(1-a_*)\,.
\eeqa
Then $\E$ and ${\cal L}$ are given by
\beqa
\E&=&\frac{\sqrt{3+4b^2}-b}{3}+
\frac{2^{2/3}(\sqrt{3+4b^2}-b)^2}{3\sqrt{3+4b^2}}(1-a_*)^{1/3}
-\frac{45-7b^4+4b(3+4b^2)^{3/2}}{2^{5/3}3(3+4b^2)^{3/2}}(1-a_*)^{2/3}+
O(1-a_*)
\\
{\cal L}/M&=&\frac{2(\sqrt{3+4b^2}-b)}{3}+
\frac{2^{5/3}(\sqrt{3+4b^2}-b)^2}{3\sqrt{3+4b^2}}(1-a_*)^{1/3}
\nonumber\\ &&
+\frac{9+2b\left(72b+86b^3-5(3+4b^2)^{3/2}\right)}{2^{2/3}3(3+4b^2)^{3/2}}(1-a_*)^{2/3}+
O(1-a_*)
\eeqa

On the other hand, for a Schwarzschild black hole, the ISCO radius for a prograde motion 
for $b\rightarrow\infty$  is given by 
\beqa
r_I^{a_*=0}/M=2\,.
\label{isco:l>0:a=0}
\eeqa
For a slowly rotating black hole with large $b$ ($1\gg 1/b\gg a_*$), 
the correction to  Eq.~(\ref{isco:l>0:a=0}) is~\cite{igata}
\beqa
r_I/M-r_I^{a_*=0}/M= 
\frac{2}{\sqrt{3}b}-\frac{8}{9b^2}+\left(-\frac{2}{3^{1/4}{b}^{1/2}}
+O(b^{-3/2})\right)a_* + O(b^{-3})
+O(a_*^2) \,,
\eeqa
and  $\E$ and ${\cal L}$ becomes
\beqa
\E&=&\frac{2}{3^{3/4}{b}^{1/2}}+\left(\frac{b}{2}+O(b^0)\right)a_*+O(b^{-3/2})+O(a_*^2)\\
{\cal L}/M&=&2b+2\sqrt{3}+(-2(3^{3/4}) b^{1/2}+O(b^{-1/2}))a_*+O(b^{-1})+O(a_*^2)
\eeqa

\subsection{ISCO for retrograde motion}
We seek
a solution to Eq.~(\ref{risco}) 
for a retrograde motion (${\cal  L}<0$). 
As found by Aliev and Ozdemir~\cite{aliev}, in the case a maximally rotating black hole ($a_*=1$), 
the radius of the ISCO for a retrograde motion for $b\rightarrow\infty$  is given by
\beqa
r_I^{a_*=1}/M=2 + 4\,\cos \left(\frac13\arctan \left(\frac{\sqrt{7}}{3}\right)\right)
\simeq 5.884 \,.
\label{isco:l<0}
\eeqa
For a nearly maximally rotating and a large $b$ (specifically we consider 
the case ${ O}(1/b^2)\sim {O}(1-a_*)$), the correction to  Eq.~(\ref{isco:l<0}) is
\beqa
&&r_I/M-r_I^{a_*=1}/M=\alpha(1-a_*)+\beta b^{-2}+ {{O}(1-a_*)^2}+ 
    {O}(b^{-3})\,\\
\alpha&=&\frac{8\,\left( 4466012 + {\sqrt{79916110092053}}\,
       \cos \left(
\frac13
\arctan \left(\frac{2870281010308837411\,{\sqrt{7}}}{5715327328333426410209}\right)
\right) \right) }{-25156453 + 
    4\,{\sqrt{158178564944911}}\,\cos \left(
\frac{2\,\pi}{3}  +
\frac13 
         \arccos \left(\frac{-324799494986675497691\,{\sqrt{\frac{7}{22596937849273}}}}{180775502794184}\right)
\right)}\nonumber\\
&\simeq& -1.42114\,\nonumber\\
\beta&=&\frac{-22\,\left( 56057 + 6\,{\sqrt{298562594}}\,\cos \left(
\frac13
\arctan \left(\frac{2220413061871\,{\sqrt{7}}}{40850566758917}\right)
\right) \right) }{-75469359 + 12\,{\sqrt{158178564944911}}\,
     \cos \left(
\frac{2\,\pi}{3}  + 
\frac13
\arccos \left(\frac{-324799494986675497691\,{\sqrt{\frac{7}{22596937849273}}}}{180775502794184}\right)
\right)}\nonumber\\
&\simeq& 1.55107\times 10^{-2}\,.
\eeqa
The asymptotic form of  $\E$ and ${\cal L}$ is given by
\beqa
\E&=&\left( {7.82673}{{b}} + 8.72195\times 10^{-2} b^{-1}    \right)  + 
  \left( -{2.07312}{b} + 1.80836\times 10^{-2} b^{-1}  \right) \,(1-a_*) 
\nonumber\\ &&
+ {{O}(1-a_*)^2}+ 
    {O}(b^{-3})\,\\
{\cal L}/M&=&\left( {-42.614}{b} - 0.37131\, b^{-1}   \right)  + 
  \left( {20.0393}{b} + 5.03812\times 10^{-3} b^{-1}    \right) \,(1-a_*) 
+ {{O}(1-a_*)^2}+ 
    {O}(b^{-3}) \,,
\eeqa
which seems to agree with Eq.(43) in~\cite{aliev} although the limiting value of ${\cal L}$ slightly deviates from theirs. 

On the other hand, for a Schwarzschild black hole, the ISCO radius for a retrograde motion 
for $b\rightarrow\infty$  is given by 
\beqa
r_I^{a_*=0}/M=\frac{5+\sqrt{13}}{2}\simeq 4.30278\,.
\label{isco:l<0:a=0}
\eeqa
For a slowly rotating black hole with large $b$ ($1\gg 1/b\gg a_*$), 
the correction to  Eq.~(\ref{isco:l<0:a=0}) is
\beqa
r_I/M-r_I^{a_*=0}/M= \sqrt{\frac{107+41 \sqrt{13}}{78}} a_*+
\frac{1}{234}\left(41 \sqrt{13}-143\right) {b}^{-2} + O(b^{-3})+O(a_*^2) \,.
\eeqa
Then $\E$ and ${\cal L}$ take the forms 
\beqa
\E&=& \left(\sqrt{\frac{1}{3} \left(46+13 \sqrt{13}\right)}b+
\frac{1}{6b} \sqrt{\frac{1}{3} \left(\sqrt{13}-2\right)}
\right)
+ \left(\frac{4+\sqrt{13}}{3}b+\frac{\left(-5019+1250 \sqrt{13}\right)
   }{11934b}\right)a_*
\nonumber\\
&&
+O(b^{-3})+O\left(a_*^2\right),\\
{\cal L}/M &=& \left(-\frac{47+13\sqrt{13}}{4}b+\frac{1-\sqrt{13}}{6b}\right)
+
\left(-\sqrt{\frac{1}{6} \left(1013+281 \sqrt{13}\right)} b+\frac{81}{13b}\sqrt{\frac{6}{439183+121829 \sqrt{13}}}\right) a_* \nonumber\\
&&+O(b^{-3})+O(a_*^2),
\eeqa
which coincide with the results by Frolov and Schoom~\cite{fs} in the limit of $a_*\rightarrow 0$.



\begin{thebibliography}{10}

\bibitem{rees}
M.~J.~Rees,
  Ann.\ Rev.\ Astron.\ Astrophys.\  {\bf 22}, 471 (1984).
  
\bibitem{ss}
N.~I.~Shakura and R.~A.~Sunyaev,
  Astron.\ Astrophys.\  {\bf 24}, 337 (1973).
  
\bibitem{bz}
R.~D.~Blandford and R.~L.~Znajek,
  Mon.\ Not.\ Roy.\ Astron.\ Soc.\  {\bf 179}, 433 (1977).

\bibitem{tnm}
A.~Tchekhovskoy, R.~Narayan and J.~C.~McKinney,
  Mon.\ Not.\ Roy.\ Astron.\ Soc.\  {\bf 418}, L79 (2011).

\bibitem{tnm2}
 A.~Tchekhovskoy, R.~Narayan and J.~C.~McKinney,
  Astrophys.\ J.\  {\bf 711}, 50 (2010)
  [arXiv:0911.2228 [astro-ph.HE]].


\bibitem{fabian}
A.~C.~Fabian, M.~J.~Rees, L.~Stella and N.~E.~White,
  Mon.\ Not.\ Roy.\ Astron.\ Soc.\  {\bf 238}, 729 (1989).

\bibitem{reynolds}
C.~S.~Reynolds,
  arXiv:1302.3260 [astro-ph.HE].

\bibitem{zcc}
S.~N.~Zhang, W.~Cui and W.~Chen,
  Astrophys.\ J.\  {\bf 482}, L155 (1997)
  
\bibitem{mn}
J.~E.~McClintock, R.~Narayan and J.~F.~Steiner,
  arXiv:1303.1583 [astro-ph.HE].


\bibitem{aliev}
A.N. Aliev and N. Ozdemir, 
Mon.\ Not.\ Roy.\ Astron.\ Soc.\  {\bf 336}, 241 (2002).

\bibitem{igata}
T. Igata, T. Harada and M. Kimura, 
Phys.\ Rev.\ D {\bf 85}, 104028 (2012).

\bibitem{Prasanna:1978vh} 
  A.~R.~Prasanna and C.~V.~Vishveshwara,
  Pramana {\bf 11}, 359 (1978).
  
\bibitem{fs}
 V.~P.~Frolov and A.~A.~Shoom,
  Phys.\ Rev.\ D {\bf 82}, 084034 (2010)
  [arXiv:1008.2985 [gr-qc]].

\bibitem{zfs}
A.~M.~A.~Zahrani, V.~P.~Frolov and A.~A.~Shoom,
  Phys.\ Rev.\ D {\bf 87}, no. 8, 084043 (2013)
  [arXiv:1301.4633 [gr-qc]].

\bibitem{Hussain:2014cba} 
  S.~Hussain, I.~Hussain and M.~Jamil,
  arXiv:1402.2731 [gr-qc].


\bibitem{Preti:2009zz} 
  G.~Preti,
  Int.\ J.\ Mod.\ Phys.\ D {\bf 18}, 529 (2009).


\bibitem{Preti:2010zza} 
  G.~Preti,
  Phys.\ Rev.\ D {\bf 81}, 024008 (2010).


\bibitem{frolov}
V.~P.~Frolov,
  Phys.\ Rev.\ D {\bf 85}, 024020 (2012)
  [arXiv:1110.6274 [gr-qc]].


\bibitem{zahrani}
A.~M.~A.~Zahrani,
  arXiv:1407.7069 [gr-qc].

\bibitem{gnedin}
M.~Y.~Piotrovich, Y.~N.~Gnedin, S.~D.~Buliga, T.~M.~Natsvlishvili, N.~A.~Silant'ev and A.~S.~Nikitenko,
  arXiv:1409.2283 [astro-ph.SR].

\bibitem{wald}
R.M. Wald, 
Phys.\ Rev.\ D {\bf 10}, 1680 (1974).
  

\bibitem{klk}
A.~R.~King, J.~P.~Lasota and W.~Kundt,
  Phys.\ Rev.\ D {\bf 12}, 3037 (1975); 
J. Bicak and V. Janis, Mon.\ Not.\ Roy.\ Astron.\ Soc.\  {\bf 212}, 899 (1985).






\end{thebibliography}
\end{document}